IHF Bayerisches Staatsinstitut für Hochschulforschung und Hochschulplanung

# Leadership, Cooperation and Conflicts in Physics: Research Leaders' Perspectives

Maike Reimer, Bianca Burkert, Theresa Görg, Isabell M. Welpe

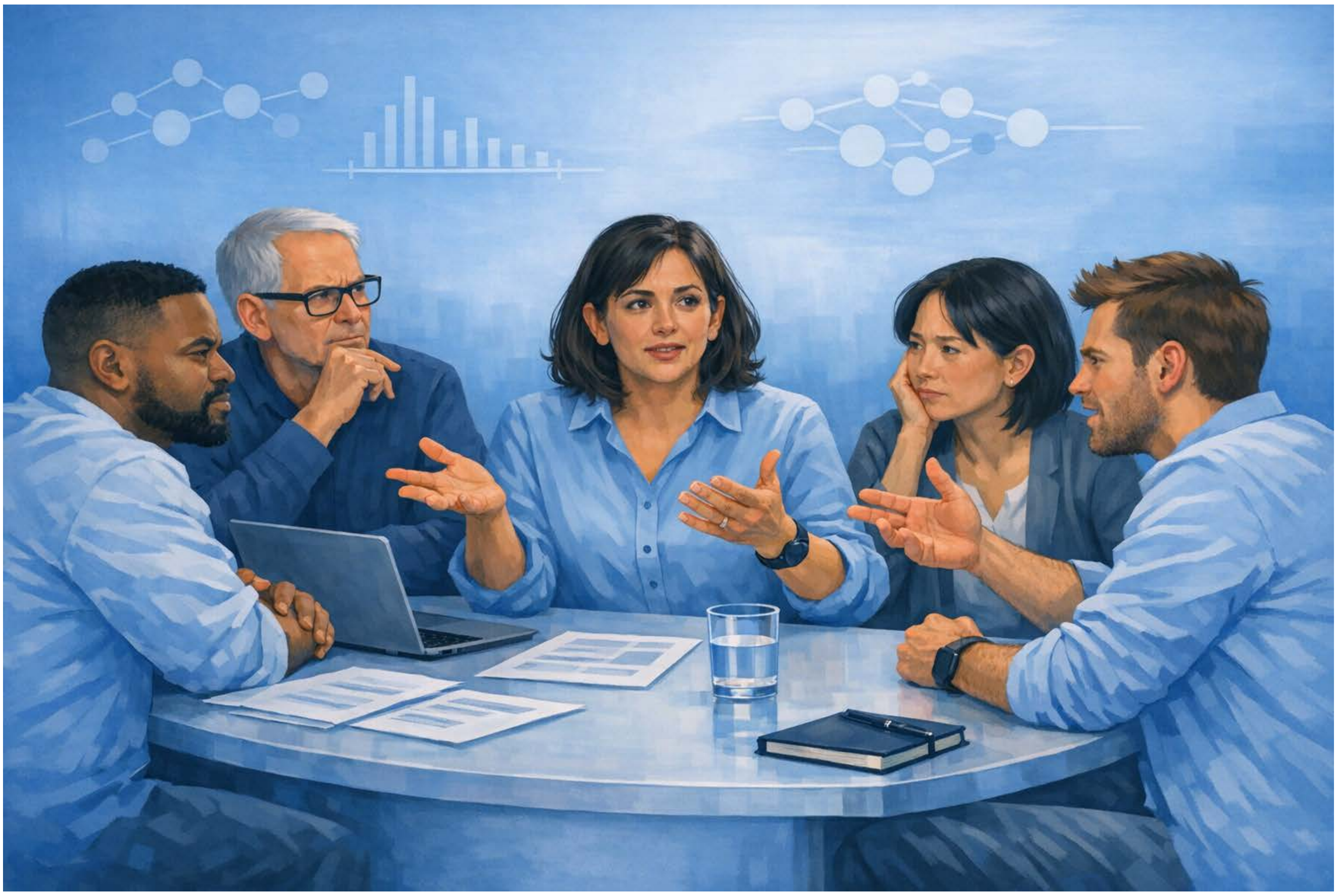

# Imprint

## Authors

Maike Reimer (IHF), Bianca Burkert (IHF), Theresa Görg (IHF), Isabell M. Welpe (IHF/TU München)

Author contributions[1]:

Dr. Maike Reimer: Conceptualization, Methodology, Validation, Formal Analysis, Data curation, Writing Original Draft, Writing – Review & Editing, Visualization, Supervision, Project Administration

Bianca Burkert: Formal Analysis, Data curation, Writing – Review & Editing

Theresa Görg: Validation, Formal Analysis, Data Curation, Writing – Review & Editing, Visualization

Professor Dr. Isabell M. Welpe: Original Idea of Study, Conceptualization, Writing – Review & Editing

## Acknowledgements

We would like to thank the following persons:

Professor Dr. Christiane Koch, Professor for Theoretical Physics at FU Berlin, and former Spokesperson of the Quantum Optics and Photonics division at DPG, for instigating this project and supporting it theoretically, practically and morally through its conception, implementation and dissemination of results, and for providing access to the insights of her colleagues by establishing contact both with interviewees and survey respondents

Frau *Oberstudienrätin* Agnes Sandner, Spokesperson of the working group on equal opportunities at the DPG, for her engaged support in preparing, testing and fielding the survey and communicating its results

Dr. Esther Ostmeier for her input in the conceptual stage, the development of the methodological approach and the conduct of interviews

Noel Schmid, Lale Altun, Susanne Hübner, Dagmara Jaworucka and William Stokes for their support in data curation, formal analyses, visualization and proof-reading during their student internships

**Most of all, we are heavily indebted to 11 anonymous physicists of all subfields and genders who provided us with insightful and reflective interview material. While we cannot disclose their names for confidentiality reasons, their perspectives and experiences have been deeply integrated into this report.**

## Citation

[1] According to the Contributor Role Taxonomy (CRediT, see https://credit.niso.org/)

## Inhaltsverzeichnis

## Figures and Tables

### Figures

### Tables

## Executive Summary

- In recent years, conflicts in research cooperations and the role of leadership in academia have become a much-discussed phenomenon in the DACH[2] region, with two central aspects:
  a) individual senior researchers who abuse their formal power and/or fail to exert their leadership functions responsibly and professionally,
  b) the formal structures of governance in universities and research organizations which are not able to deal with these problems in a fair and transparent manner.

  As most empirical insights come from junior researchers, this report focuses on research leaders' perspectives on workplace conflicts and evaluates the encompassing institutional reactions.
- About 12 % of respondents had experienced complaints, most of them concerning non-scientific misconduct (e.g. power abuse, discrimination or intimidation). About half reported that constructive attempts were made to find a fair solution to the conflict. Standards of fair dealings – such as hearing anyone involved, clearly explaining which rules or standards had been violated or providing all necessary information to the accused – were not observed in a large part of cases.
- Conflict in research teams was a near-ubiquitous phenomenon, with the three top issues being lack of respect or overconfidence, non-collegial behaviour and authorship. Most frequently involved (and perceived as most helpful) were informal sources of support, such as colleagues at the same institution and private contacts. Official institutional bodies were less often involved and often not perceived as helpful.
- In the majority of conflicts, there was no serious harm done to the research leaders involved, and qualification goals of conflicting parties could be reached. More widespread however were damages to research productivity such as delays or unpublished results. About two third of conflicts involved at least one person in a qualification process, demonstrating how inextricably research is linked with qualification, and that conflicts often occur in the complex entanglement of collaborative knowledge production and certification of individual research performance.
- Satisfaction with conflict development and its final resolution was fairly evenly distributed over the spectrum from complete dissatisfaction to complete satisfaction. Most research leaders changed their leadership practices in response to conflict experiences, showing that conflicts can be an opportunity to learn and grow.
- Within Physics as a research culture, some *field-specific differences* point to subtle variations in the contexts for the development of conflicts, partly connected to different group sizes, although overall, similarities are high. *Gender differences* are scarce; they were most pronounced and conclusive in emotional and practical responses.

[2] This acronym stands for the official initials for the three (also) German-speaking countries Germany (D), Austria (A) and Switzerland (CH).

- To improve the situation, universities and research organizations can follow five lines of action that should be implemented with respect for and participation of research professionals:
    - Normalize conflict and de-problematize its widespread existence
    - Develop research leaders' professional identity and skill set, with an emphasis on conflict management
    - Develop awareness of challenges in collaboration among followers and prepare them for how to deal with them
    - Support the administration in working together with researchers
    - Improve and professionalize organizational procedures for preventing and dealing with conflicts

# 1. Conflicts in research teams: Why leaders' perspectives?

## 1.1 Starting point: A public discussion on leadership perception

In recent years, conflicts in research cooperations and the role of leadership in academia have become a much-discussed phenomenon in the DACH[3] region with increasing attention of research, practice and media. Many contributions focus on the perspective of junior scholars (e.g. Agarwala & Scholz, 2017; Gewin, 2025; Haug, 2018). They paint a bleak picture of research leadership and organizational conflict management in Germany and its German-speaking neighbouring states, and the devastating effect especially on researchers not (yet) in tenured senior positions.

Two recurring problem areas are pointed out in the discourse:

a) individual senior researchers that abuse their formal power and/or fail to exert their leadership functions responsibly and professionally,

b) the formal structures of governance in universities and research organizations which are not able to deal with these problems in a fair and transparent manner.

In this report, in order to contribute to an improvement of the situation we address leadership (by individual research leaders) and governance (by and in institutions) in the case of arising conflicts in research teams.

While leadership in research has been explored conceptually and empirically in recent years (e.g Braun et al., 2016; Schmidt et al. 2017; Verbree et al., 2011), no quantitative or qualitative studies explore specifically the perspectives of leading researchers on conflicts within their research team and the organizational support systems available to them. Most empirical studies focus on juniors and show a substantial amount of perceived conflict, substandard leadership and supervisory behaviour on one hand, and a lack of transparent and effective ways to address these problems in their research organizations on the other hand (e.g. Ambrasat et al., 2024; Arcudi et al., 2019; Lin et al., 2024; Majev et al., 2021; Olsthoorn et al., 2020; Russell et al., 2022; Stahl, 2024; Striebing et al., 2024; Vieira Mourato et al., 2023).[4] While there is some survey evidence that professors or other senior research leaders are not wholeheartedly satisfied with procedures and fairness at their institutions either (e.g. Schraudner et al., 2019; Staub, 2020), we consider a better understanding of their perspective to be of utmost importance for two reasons:

---

[3] This acronym stands for the official initials for the three (also) German-speaking countries Germany (D), Austria (A) and Switzerland (CH).

[4] In some empirical inquiries, seniors are included but receive identical questions, and their responses are not explored separately or contrasted and discussed in comparison to those of juniors.

1. *Research leaders also suffer from badly managed conflicts.*

- In highly interdependent research groups conflicts with juniors can cause direct or indirect damage to research or seniors personally (Abou Hamdan et al., 2021; DPG, 2019; DGPs, 2022; Herschberg et al., 2018; Prevost & Hunt, 2018). Depending on their position in the research group and the relation of their work package to the overall project goals, their influence on the groups and the seniors' own research success can be substantial.
- While solid data is not available, there are some highly publicized cases in which established scientists accused of leadership misconduct appear to have been treated unfairly and without due processes by organizational bodies (e.g. Aeschlimann et al., 2019; Goebel, 2022; Schauer, 2019). Non-majority senior staff who do not conform to (implicit) standards of their research communities or organizations are more vulnerable to such experiences, both in conflicts with superiors and with subordinates - also or even more at higher hierarchy levels (e.g. Ambrasat & Heger, 2020; Armstrong, 2012; Gorlewski et al., 2014; Niemann et al., 2020; Prevost & Hunt, 2018; Staub, 2020).

*2. Research leaders are indispensable for improvement.*

- While all members of research groups can contribute to conflict prevention and resolution (e.g. Machovcova et al., 2023; May et al., 2014), responsibility for and capability of doing so falls disproportionally on the leader(s) of the group. Leaders' formal role makes them responsible for the managing of all kind of workplace dynamics, therefore, conflict management is a central leadership task and function (Zhao et al., 2018; Gelfand et al., 2012). In addition, the way leaders handle conflicts is linked to the overall conflict management culture of their organization (Gelfand et al., 2012). Consequently, leaders offer the most leverage at both the group and organizational levels and are therefore indispensable as allies for change.

We focus on conflict, since the behaviours that have been cited as examples of illegitimate use of formal power by leading researchers encompass a wide range. While some undoubtedly can be classified as ruthless and exploitative (if not downright illegal), many more instances are more ambiguous. Some of leader's behaviour may be merely misguided or unprofessional (Fehrenbach, 2020; Schmidt & Richter, 2008). Possibly, misunderstandings and failed communication, or even values and expectations that differ between fields, genders, positions or individuals are at the bottom. Similar to what has been shown about scientific misconduct, incidences in the greyish area are probably more common than clear-cut ones (Schmidt & Richter, 2008; Fehrenbach, 2020). Addressing the whole spectrum of phenomena as *conflicts* neither implies nor precludes that one side is more to blame than the other, that someone has done something reprehensible or violated established rules and standards (Baillien et al., 2017). It does however allow for a broader and more nuanced perspective on humans interacting under specific circumstances, and offers a better point of leverage for both research into and improvements of individual and organizational behaviours.

This report contributes to the understanding and prevention/management of conflict escalation in research groups and development of potentially useful suggestions for improvement by giving the leaders an opportunity to describe their perspectives on workplace conflicts and evaluate the encompassing institutional reactions. We start in *chapter 1* with a description of our conceptual background regarding leadership, conflict and institutional support systems in case of conflicts in, and the specific collaboration environment associated with the field of Physics. In *chapter 2*, a short description of data and methods and the composition of our participants follows. *Chapter 3* details our results regarding experiences with complaints/reproaches and their institutional handling. *Chapter 4* centres on a specific conflict that the participants deem noteworthy, exploring the specific circumstances and consequences. In *chapter 5*, we will sum up and suggest lines of action based on the findings and research literature.

## 1.2 Concepts, background and empirical approach

### 1.2.1 Leadership and institutional governance of conflicts

At the *individual level*, we build on *concepts and literature of leadership of research teams* (e.g. Becker, 2022; Braun et al., 2016; Verbree et al., 2011) by professors or other group leaders. Not only are they supposed to organize all staff working in their teams, they are also formally responsible for them as their superordinates (*Vorgesetzte*), with all rights and duties this entails (Becker, 2022). Ideally, team leaders have a professional approach to this task, and leadership functions are an integral part of their professional role and self-concept (Rehbock et al., 2023). Increasing team sizes and structural changes in higher education make effective leadership of research teams both more difficult and more necessary (Bronner & Frohnen, 2018; Gorlewski et al., 2014; Schmid et al., 2017), but professors and senior researchers don't always see themselves primarily as leaders of teams (Rehbock et al., 2023). Leading others is frequently perceived as an additional duty they have to perform in order to get their actual work (research) done (Rehbock et al., 2023; Schmidt & Richter, 2008; 2009) or even judged to be at odds with how research and collaboration among researchers should work (Baitsch, 2017; Lohaus, 2019). Also, professional leadership is not purposefully incentivized by research institutions, and professors are not prepared for leadership roles and tasks in their scientific careers where success is almost exclusively evaluated in terms of research proficiency (Armstrong, 2021; Schmidt & Richter, 2008; 2009).

At the level of *institutional or organizational bodies'* reactions to and management of conflict, we look at the university or research organization as an employer that organizes the activities of their staff according to principles of governance, having both authorities and responsibilities over its members. Governance encompasses the processes and structures for decision making and allocation of authority within the organization (de Boer et al., 2007). Through their governance, universities are responsible for creating a fair work environment for everyone and an organizational culture in which constructive collaboration is possible and rewarded (May et al., 2014). They play an important role for a productive and healthy working atmosphere by establishing rules, procedures and points of contact for dealing with conflicts in an appropriate manner as early as possible (Herrmann, 2021; Hoormann & Matheis, 2012; Leidenfrost & Rothwangl, 2019; Reimer & Welpe, 2021). As conflict has been recognized as a major impediment to work culture and productivity, it is recommended to implement organization-wide conflict management systems (Hochmuth, 2014; Symanski, 2012; von Knobelsdorff, 2025).

Professors or (tenured) research leaders have a comparatively high degree of autonomy in choosing their work goals and content, designing their work organization and selecting, incentivizing and managing the staff within their research teams (Bronner & Frohnen, 2018). At the same time, senior researchers exert substantial governing power themselves via mechanisms of collegial self-government (Banscherus, 2018). Administrative structures, however, follow a formal bureaucratic logic of public administration (Banscherus, 2018; Nickel, 2012). The top-down establishment of central rules and structures is a challenging task in such an environment and often met with scepticism and open or covert resistance by scientific staff (Baitsch, 2017).

### 1.2.2 Physics as research culture

Physics presents a specific environment in which the modes of collaboration, processes of knowledge generation and the nature of research itself shape occurring conflicts. Physics is a highly interdependent team science carried out mainly in research groups (as the basic unit of knowledge production) that range from smaller ones (about 5 – 10) to large consortia of several hundred members. Research groups typically consist of a three-tier pyramidal structure, with the professor or otherwise most senior (often tenured) person at the top as principal investigator, several postdocs at the middle level, and a higher number of persons pursuing a PhD at the bottom. Postdocs often have pre-doctoral researchers assigned to work with them on circumscribed research avenues, so that in the day-to-day work they are their leaders and supervisors of their theses, while the professor or similar remains the formal supervisor and superordinate. Research groups, therefore, consist in great part of ECRs as apprentices who are being trained in practical research and socialized into a professional identity of a colleague (DPG, 2019; Gläser & Laudel, 2015; Laudel & Gläser, 2008). International diversity is high, with English being not only the language of most publications but also of daily collaboration.

Research in Physics is highly cooperative, collective and cumulative, based on principles of specialization and labour division (Knorr-Cetina, 2002). Pre-doctoral researchers and even persons pursuing their master’s degrees are fully included in the collaborative enterprise for the duration of their qualification phase. Individual contribution or merit is hard to distinguish unequivocally from outside, and authorship of manuscripts in high-ranking journals or at prestigious venues becomes an indispensable signal of individual contribution to research outcomes (Frisch et al., 2022; Knorr-Cetina, 1999; Galison, 2013; Sorgner, 2022). Publication rate is comparatively high and author lists can be long (Johann et al., 2020; Sorgner, 2022), often including contributions of researchers who have left the organization already at the time of final release. As pre- and postdoctoral researchers need clear signals of their research productivity in a limited time frame, and seniors establish their standing among their peers and their funding-worthiness mainly via high-ranking publications, authorship can be serious business (DPG, 2019; Frisch et al., 2022).

While limiting generalizability, the concentration on one field of research allows us to eliminate field-specific heterogeneity that may obscure patterns found within fields. At the same time,

Physics does provide a level of internal differentiation into subfields that can be meaningfully explored.

## 2. Methods and data[5]

We conducted eleven interviews with professors in Physics in the DACH area who were recruited through personal contacts and further snowballing. Due to the highly sensitive nature of the topic and the limited possibilities for our research team to contact interviewees, there was no selection; we included all persons who volunteered to share their views on leadership and conflict experiences. Interviews were transcribed, anonymized and systematically explored for recurring themes relating to leadership of research groups with respect to conflicts and the institutional support received.

Based on the identified patterns and themes, an online survey was developed, tested and conducted between October and December 2024 among professors and research group leaders in Physics. These were invited via the mailing list of the Deutsche Physikalische Gesellschaft (DPG). After conducting descriptive quantitative analyses, interviews and free texts were mined for illustrative quotes to enrich and contextualize the abstract and preformatted responses.

Empirically, we focus a) on conflicts where leaders were confronted with complaints about their behaviours, and b) on selected conflicts that respondents chose from their experiences[6]. We assess characteristics of the persons involved, and the consequences both for the research work and the leaders' individual person. In addition, we look at experiences with institutional and other support, subsequent changes of leadership practices, and satisfaction with conflict management and resolution. We focus on only own experiences of conflict, and a single experience per person. The common practice in surveys to ask about experienced and witnessed events, and assess further details about the characteristics of all experiences combined, has the advantage of covering all relevant instances, but does so at the cost of obscuring differences between single experiences.

After data cleansing, 385 respondents remained in the data set. Due to participants with limited conflict experiences, dropouts and missing values on some variables, this number is lower for most of the analyses shown. The majority of respondents (75.7 %) gave their gender as male[7], and almost 90 % were German nationals (table 7). They were predominantly working at universities (59.8 %), while non-university research organizations made up another 27 % combined, with the Helmholtz association contributing the largest share, followed by Max-Planck-

---

[5] For a full method report, see Reimer et al., 2026

[6] Respondents were asked to select "*if possible [a conflict] that you have found particularly serious, important or typical - that may have been particularly challenging, troubling or noteworthy for you.*" We therefore expect a sample of the more serious conflicts for each respondent, however, with a wide range of how serious these were in the absolute.

[7] A category "diverse" was included, but due to the low numbers of persons selecting it, the category is not displayed both for methodological and privacy reasons.

Society, Leibniz Association and Fraunhofer Institutes. 13.2 % reported other institutions[8]. A bit over half (53.8 %) were professors, followed by 38.3 % post-doctoral researchers or junior research group leaders, with “other” making up 7.9 %[9].

**Figure 1:** *Participants’ characteristics: gender, subfield, institution type and position*

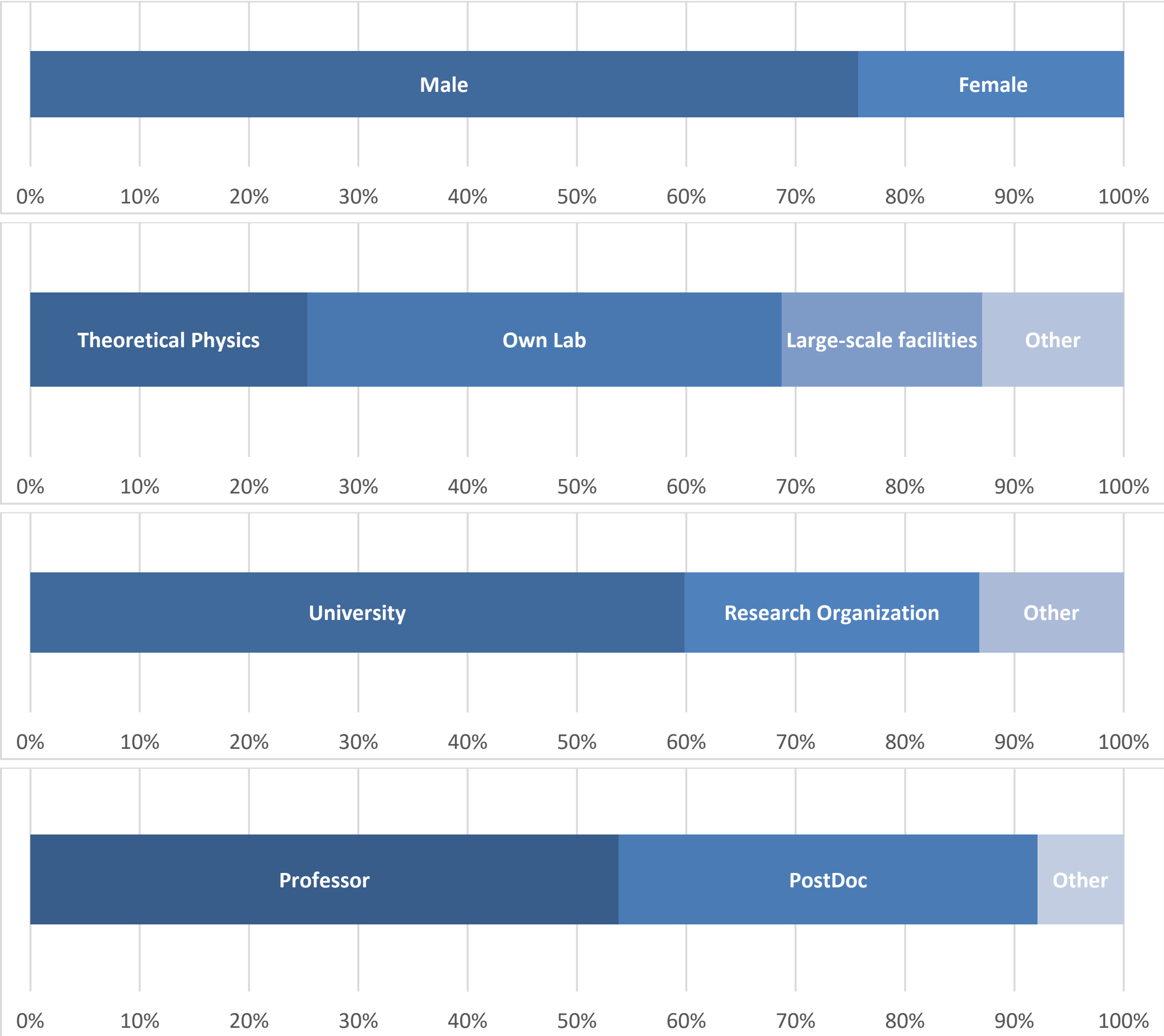

Source: DPG Survey 2024, own analyses, N= 280/368/371/371

The largest group (43.4 %) were experimental physicists with their own lab, followed with a margin by theoretical physicists (25.3 %) and experimental physicists doing research at large-scale facilities (18.3 %, figure 1). The remaining 12.9 % reported areas such as “physical chemistry”, “didactics” or “applied research”.

[8] Therse included universities of applied sciences, organizations in other countries that could not be placed within the categories offered, private companies, governmental institutions and persons in retirement.

[9] Persons who are not in an active research leadership position at the time of responding are nonetheless included, since reports can be made about conflict experiences in any of their past leadership positions.

There is no independent and valid information on the survey target group against which the samples' representativeness could be assessed. In particular, we have no way to estimate whether persons with specific patterns of conflict experiences have elected to participate at a higher or lower rate, so that generalizations regarding the level and kind of conflicts must be undertaken with caution. For an attempt at estimating the sample's biases with respect to central demographics, see Reimer et al. (2026).

# 3. When things go south: Experiences with complaints

## 3.1 Prevalence and kind of complaints

50 of the surveyed researchers (11.8 %) had personally experienced one (or more) complaint(s) by their group members – either directly to themselves or to other persons, including institutional actors such as ombuds offices, staff representatives or deans. Of those, 15 (4 %) concerned scientific misconduct (such as plagiarism, falsification of data or inappropriate author attributions), and 6 (2 %) "other" reasons[10]. The majority (11 %) gave non-scientific misconduct as the reason for the complaints, such as power abuse, discrimination or intimidation (figure 2). The accusations were made about equally often by postdocs, predocs and others. These numbers are unfortunately too low to unveil differences between subgroups (such as field, gender, age/seniority or institution type) or reliably identify patterns in antecedents, circumstances and consequences.

**Figure 2:** *Research leaders' experiences of complaints*

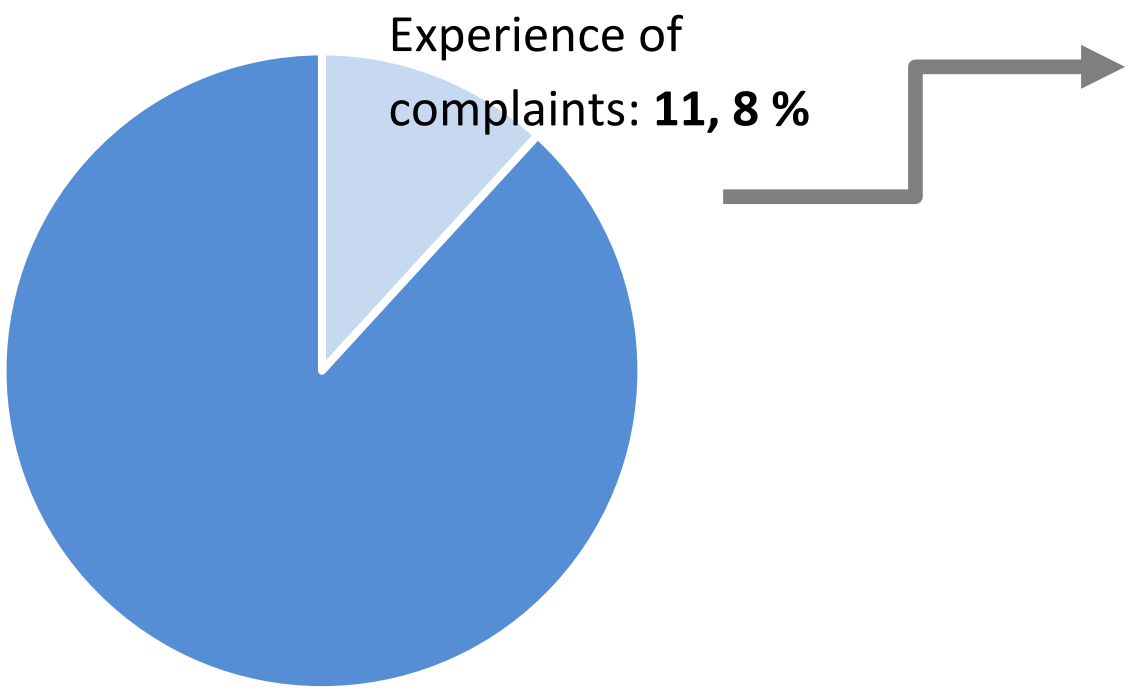

**4% Scientific misconduct** (e.g. plagiarism, falsification of data, author attribution or sequence ...)
**11% Non-scientific or leadership misconduct** (e.g. power abuse, discrimination, intimidation ...)
**2% Other kinds of misconduct** (e.g. lack of loyalty, delay of publications, unfair assessment)

Source: DPG Survey 2024, own analyses, N= 272, multiple responses

## 3.2 How were complaints dealt with?

About 50 % of respondents experienced complaints to third parties. These persons were asked whether, in their view, constructive attempts to mediate had been made with the goal of finding a fair solution that all sides can live with. In addition, we asked about the degree to which the ensuing processes reflected central criteria for fair and transparent dealing, such as being adequately heard or given full information about the complaints (figure 3).

[10] Among the examples given by respondents were "lack of loyalty towards a leader", "delaying publications" or "unfair evaluation of a doctoral candidate".

**Figure 3:** *Research leaders' evaluation of conflict response by third parties*

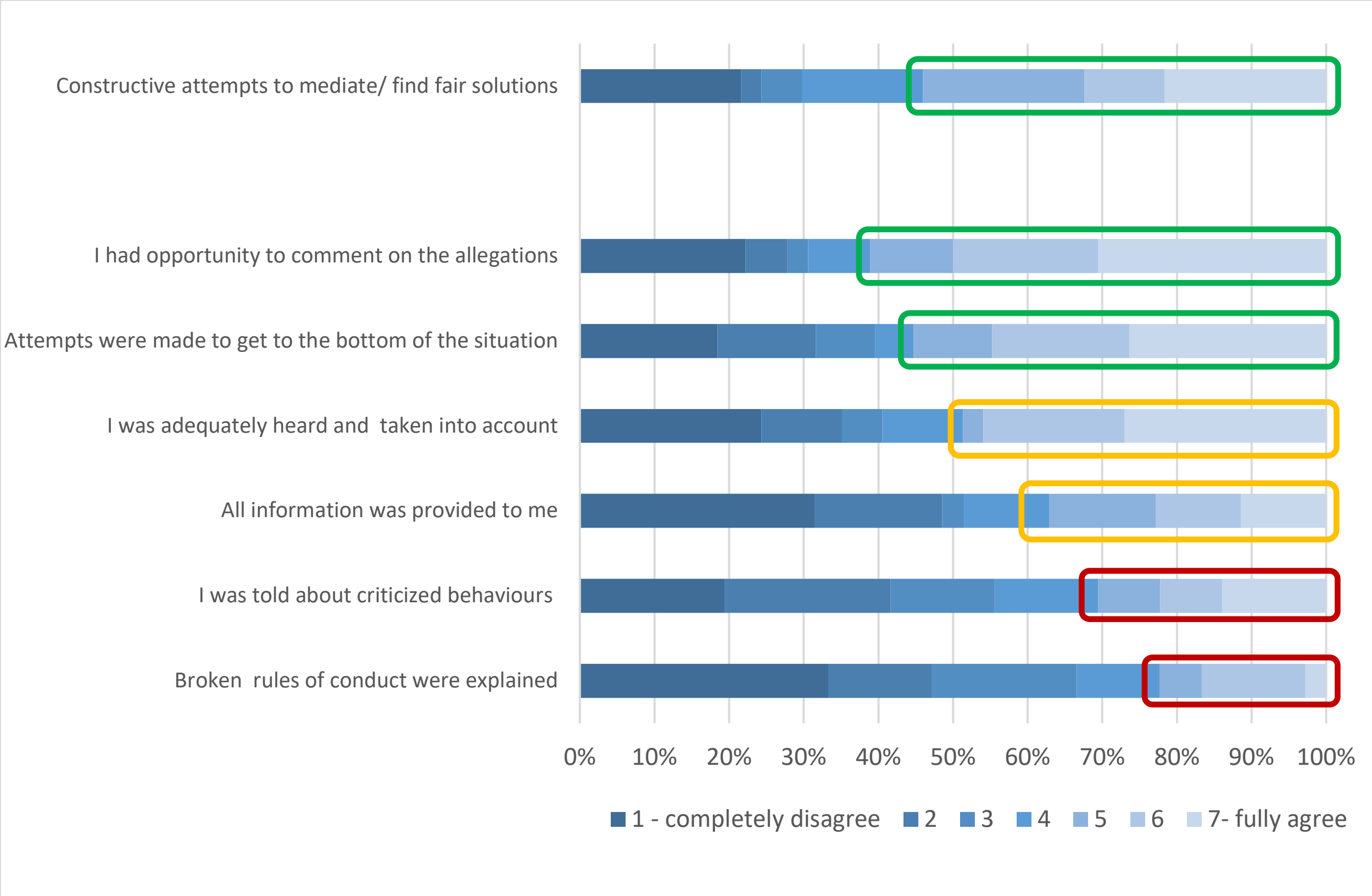

Source: DPG Survey 2024, own analyses, N= 21-23

Constructive attempts to find a fair solution to the conflict were experienced to at least some degree by a bit more than half (ratings of 5 or higher). This was also the case for two of the aspects of fair procedures: the majority agreed that attempts had been made to get to the bottom of the situation, and that they had been given an opportunity to comment on the allegations.

For the other aspects, this was not the case: less than half agreed that they received all the information that they needed to defend themselves, were adequately heard, were told in detail which of their behaviours sustained the accusation or received an explanation about which rules of conduct they had allegedly broken. The last two aspects were experienced by less than a third of respondents. Again, there is no possibility to quantitatively break these results down by field, gender, age/seniority or institution type. Table 1 shows excerpts from open-ended survey responses and interviews, contrasting cases with higher and lower ratings of fairness and mediation in the handling of accusations.

**Table 1**: *Research leaders' evaluation of conflict response – illustrating quotes*

| Positive experiences | Negative experiences |
|---|---|
| „It was about claims to leadership and the distribution of funds within the group. All researchers involved in leading the group and a mediator were present at the discussion." | „The conflict was reported directly to my supervisor without even consulting me first. Our meeting to resolve the conflict therefore came as a complete surprise to me, and I was unprepared." |
| „Some doctoral students complained that my requirements for the doctorate were too high and did not correspond with university regulations. But my boss backed me up - those were the internal rules of *[organization name]*." | „I received a call from the Human Resource manager, who informed me about the official complaint. For data protection reasons, they were not allowed to disclose any details about the person or the exact facts. [...] The result was nights of speculation about who it could have been and what the reason might have been." |
| „When things looked dire for *[the graduate student charged with plagiarism]*, they called the *[head of unit]* and accused me of discrimination and *[racism]*. The *[head of unit]* informed me about this, and that they would not pursue the accusations against me any further." | „Formally, I was not given any information about the person or the reasons for the negative evaluation. (…) I was not allowed to comment or inform the committee for promotion of the involvement of the office for conflicts. The protection of their identity took priority over my credibility." |

Source: DPG Survey/Interviews 2024. Original quotes slightly smoothed and compressed for better readability and if necessary, translated from German for the purpose of this table. For reasons of confidentiality, there are no pseudonyms or codes, and all references to genders, institution types etc. are obscured and paraphrased.

# 4. It happens every day: Experiences with conflicts

## 4.1 Prevalence and kind of conflict experience

About 93 % of the respondents have experienced one or more conflicts in one of their present or past positions as research leaders. The majority of conflicts happened in German institutions, about 60 % of cases in universities. The top issues were authorship, lack of respect and non-collegial behaviour, which are reported by about half of respondents (figure 4 and table 2). Least widespread were conflicts over access to or allocation of resources, scientific misconduct, adherence to occupational safety standards and contract renewal, with fewer than a quarter of respondents mentioning them. Experience breadth varied, too: There were respondents who reported only one and some up to 12 different conflict issues they had experienced; the median was 4.

We find no significant gender differences in the number or kind of conflict issues mentioned. With respect to the subfield, there were some identifiable patterns: Theoretical physicists reported overall fewer conflicts and less often mentioned non-collegial behaviour, issues with working hours, lack of receptivity to criticism and (unsurprisingly) adherence to occupational safety standards. Reversely, issues with contract renewals seems to be a more pronounced problem in large equipment research (not displayed).

**Figure 4:** *Conflicts in general: Most common conflict themes*

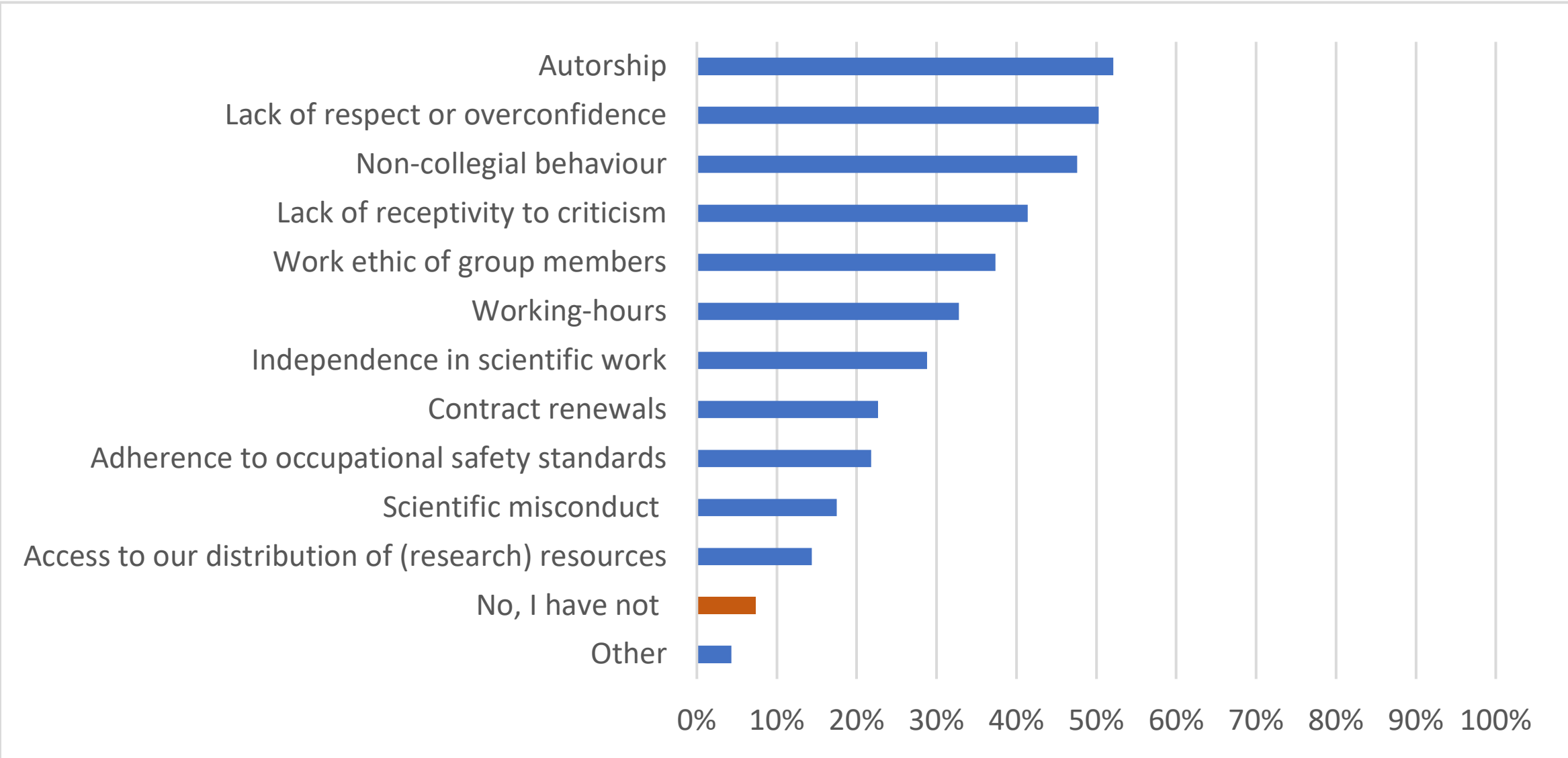

Source: DPG Survey 2024, own analyses, N= 326

**Table 2:** *Conflict themes - illustrating quotes*

| **Category** | **Illustrating Quotes** |
|---|---|
| Authorship | • „In my view, [X] shouldn't have been an author, but *[my colleagues]* warned me about the consequences if I were to insist – that [X] might go to the editor. That [X] might accuse me. Something always sticks, no matter how it turns out in the end, no matter if I'm right… the potential danger to my reputation."<br>• „Should I have been more directive in the matter? Should I have said 'You're first author, you're second, and you're last?' I had thought that they'd be able to agree among themselves, and it didn't work out at all." |
| Lack of respect or overconfidence | • „[X] was cocky – bold, arrogant somehow…. maybe trying to cover their insecurity a bit."<br>• „If [X] speaks to me as their supervisor, there is a certain hierarchy after all, and [X] talked to me in a very disrespectful manner. No one should talk to me in that way, no matter who."<br>• „My boss supported me after he had understood the problem completely (overconfidence in professional matters, immaturity, underestimation of time and effort required for scientific work, lack of receptivity to criticism)." |
| Non-collegial behavior | • „And then, [X] announced at very short notion that they're away for a month, and that was at a bad time for the project. And it was the way they said it *[mimics matter-of-fact voice]*: 'There, I'll be away. I'm gone.' "<br>• „ [X's] situation in the group was more and more untenable – the others ganged up upon them and were dissatisfied because they had to compensate for [X's] insufficient performance… Science can be quite poisonous, among professor too, it's about honour and personal slights can lead to veritable wars and everything is taken much too seriously."<br>• "It wasn't an escalated conflict, rather a long-term dissatisfaction. Surprisingly often, it is an egoistical attitude: What belongs to my doctoral thesis, and what do I *have* to do for the team." |
| Lack of receptivity | • „[X] was rather immune to good advice."<br>• „I was of the opinion that I had made it clear that I was dissatisfied, and with what, and what I expected to be done. But I don't think that [X] understood what the problem was from my point of view."<br>• „[X] started to compete with me, tried to find fault with my work, but committed grave faults themselves… Initially, I tried to solve matters by explaining, but failed due to [X's] lack of insight and ability to take criticism." |
| Work ethic / Working hours | • „[X] was very strict about sticking to an eight-and-a-half-hour day, no matter what was going on during the week. [...] When you work in a laboratory, you need to be flexible and prepared to stay longer, sometimes much longer. But that's not something I want to have to say as a group leader."<br>• „Lives function differently today than they did 20 years ago. People are more committed to their private lives. They don't say 'I'll stay until midnight and then, tomorrow, I'll come in the afternoon', but rather 'I've arranged to meet a friend at the gym at eight' or something of the kind."<br>• „It will result in tensions if a dissertation is seen as a job where the title is the result of cumulated work hours." |
| Independence | • „[X] needed constant support and supervision and, basically, never really understood why things should be done the way I wanted."<br>• „[X] was very capable in the lab with the technical, the experimental work, but when it came to analysis and interpreting results, didn't reach the necessary level and simply failed to get an independent scientific project together." |
| Contract renewals | • „After three years, [X] continued their PhD without pay, because the funds were exhausted. I had told them so in time, but they hadn't taken it seriously enough to look for funding options."<br>• „There was a gap of one and a half year between the end of one contract and the next. We tried to cover that by making the next contract full-time, but that didn't succeed immediately – and then [X] threatened, indirectly, in a veiled way, to contact a law-yer."<br>• „If a funding programme is postponed by the state, or federal state, it's hard to explain that the prospects of contract extensions aren't going to materialize due to politics." |
| Adherence to safety standards | • „A post-doc with two very high-class stipends continuously ignored safety rules."<br>• „A post-doc, after 2 years of no progress, quarrelled with a high-performing PhD and refused to adhere to safety standards." |
| Scientific misconduct | • "It was in my view a case of scientific misconduct, and I decided to write to the journal and demand retraction."<br>• „When I read the dissertation, I became suspicious that some parts were not from that person at all and took a closer look. Then it became apparent that there was extensive plagiarism from other theses from [X]'s previous working groups."<br>• „A graduate student had plagiarized in the first version of the thesis when describing the measurement setup. I told them that they would have to delete or rewrite those passages and all passages of that kind before submitting. [X] did it, too – it wasn't a big conflict, but my amazement that they weren't even aware of the problem!" |
| Resources | • „I heard that *[post-doc X]* had felt annoyed and excluded, because I gave project parts to a pre-doc that *[post-doc X*] would have liked to work on. Such tensions arise frequently between new postdocs and advanced predocs."<br>• „*[Post-doc X]* left the university, but there was this expensive device they had bought and wanted to keep access to."<br>• „It was about the degree to which post-docs are entitled to enlist others for their own work - the post-docs work, that is."<br>• „I think one contribution to this was also the layered structure of that specific grant which included multiple scientific administra-tors, who were professors higher up, and they wanted access to some of the resources, i.e. students and funding." |

Source: DPG Survey/Interviews 2024. Original quotes slightly smoothed and compressed for better readability and, if necessary, translated from German for the purpose of this table. For reasons of confidentiality, there are no pseudonyms or codes, and all references to genders, institution types etc. are obscured and paraphrased.

The respondents with conflict experiences had for the most part no difficulty in selecting a memorable conflict.[11] The conflict themes they gave for this conflict correspond in their relative frequency to the ones mentioned most frequently for conflicts in general, with some slight differences in order. The three top issues are lack of respect or overconfidence, followed by non-collegial behaviour and authorship (figure 5).

In the selected conflict themes, no gender differences appear either. Also, field differences are scarce, with the exception of authorship issues: These were reported less often as having played a part by the large equipment researchers (not displayed).

**Figure 5:** *Selected conflict experiences: Most common conflict themes*

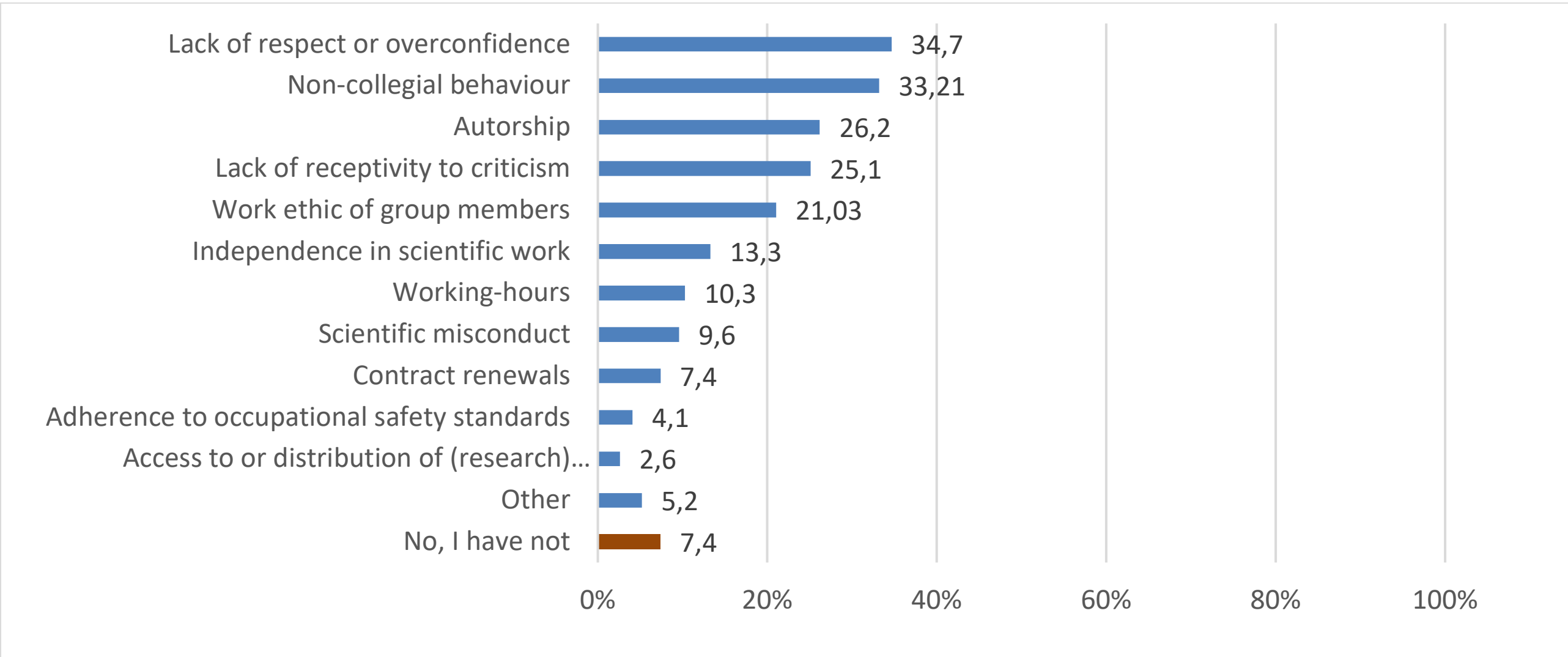

Source: DPG Survey 2024, own analyses, N= 271

## 4.2 Who has conflicts with whom?

The selected conflicts both involved the research leader directly as concerned party – in 63.1 % of cases, they indicated this at least to some degree - and additionally or exclusively in their role as team leaders in conflicts between other members of the team (figure 6). If they were involved more as conflict managers rather than as conflicting parties, they more often reported conflicts about authorship, work ethic and non-collegial behaviour (not displayed).

[11] The question wording was: *"In order to better understand the conflicts research leaders encounter, how they develop, and how they are resolved (or not), we now would like to learn more about one of these conflicts that you have experienced in your role as a leader. Of course, we will only ask general questions about the circumstances, topics and consequences and not about the specific place, time and people! Please focus on one of the conflicts you have experienced - if possible one that you have found particularly serious, important or typical - that may have been particularly challenging, troubling or noteworthy for you. It may be a conflict that was resolved satisfactorily - one that had unpleasant consequences - or one that `fizzled out´."*

**Figure 6:** *Role of research leader in the selected memorable conflict: conflicting party or conflict manager?*

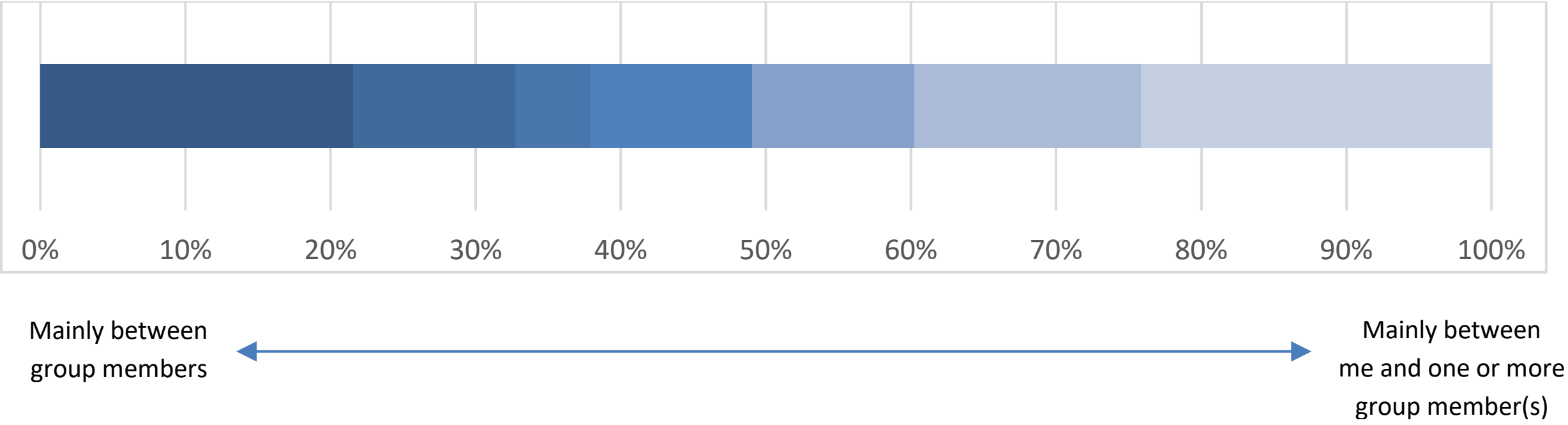

Source: DPG Survey 2024, own analyses, N= 269. *"Where did this conflict take place? More between your group members, whereby you as the leader had to take notice - or rather between you and one (or more) group member(s)?", Scale from 1 - primarily between my group members – 7-primariliy between me and my group members*

The research leaders gave their own position at the time of the conflict most often as professor (49.3 %) or postdoc/research group leader (43 %), and 8 % as "other". The main other conflicting party was most often a postdoctoral (40 % and 47.3 % of cases, Figure 7) or a predoctoral researcher (48.5 % and 34.6 %). In more than 80 % of reported conflicts, there were additional conflicting parties; most often (other) PhD students (40.5%) or (other) postdocs (31.1%, not displayed).

**Figure 7:** *Position of research leader and main other conflicting party*

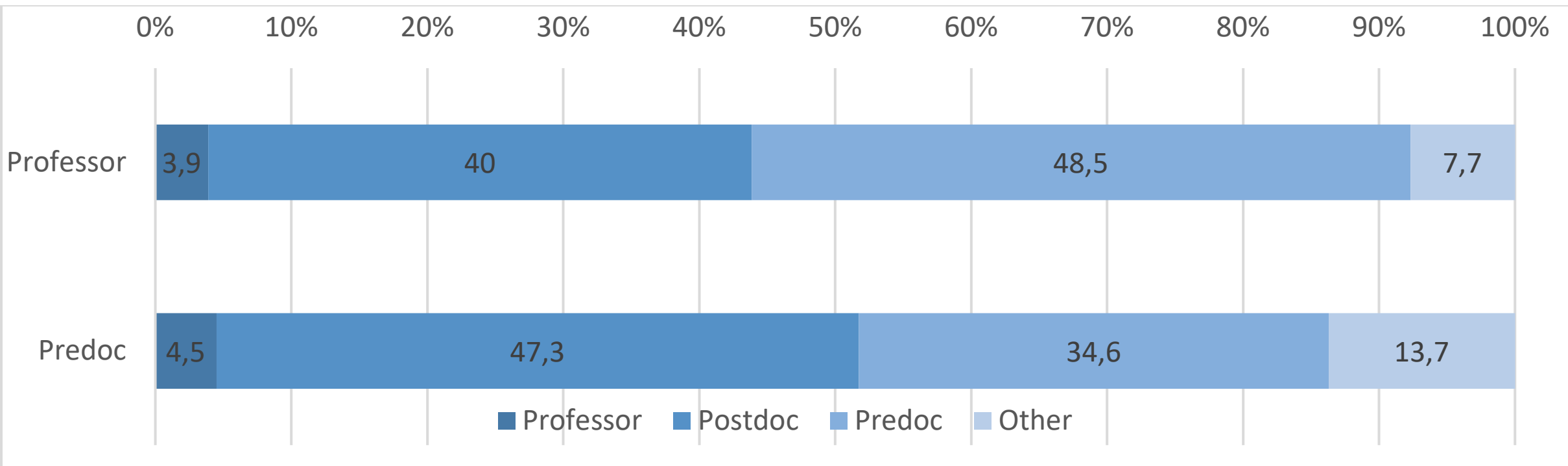

Source: DPG Survey 2024, own analyses, n=240. "Other" category for own position was left out due to small N

Conflict themes are more or less the same for all positional constellations, except for scientific independence: If the conflict occurred at the participants' post-doc level, the issue was more often about scientific independence than at the professorial stage. Moreover, the rate of independence issues was highest if the other conflicting party was a postdoc as well (not displayed).

Again, there were no observable gender differences. For the subfields, we find that at the time of the selected conflicts, theoretical physicists were more often professors than postdocs (66.7 % vs. 33.3 %), while this was almost reversed for the large appliance researchers (36.2 % vs. 63.8 %, not displayed).

## 4.3 Who is involved beyond the conflicting parties, and are they perceived as helpful?

Conflicts radiate: in addition to the conflicting parties, in about 80 % of reported conflicts, colleagues at the same institution were involved, and in 60 %, private contacts (figure 8 and table 3). Colleagues at other institutions came into play in about a third of cases. Among the institutional actors, faculty and/or university management were most frequently involved (44.7 % and 34.7 % respectively). Specialized institutional bodies followed, such as the human resource (HR) department, staff representatives, equal opportunity officers or arbitration bodies. Professional support such as mediation, coaching or external counselling came into play in about 26 %, external legal advice was sought in 12 % of cases.

**Figure 8:** *Involvement of other actors in the conflict*

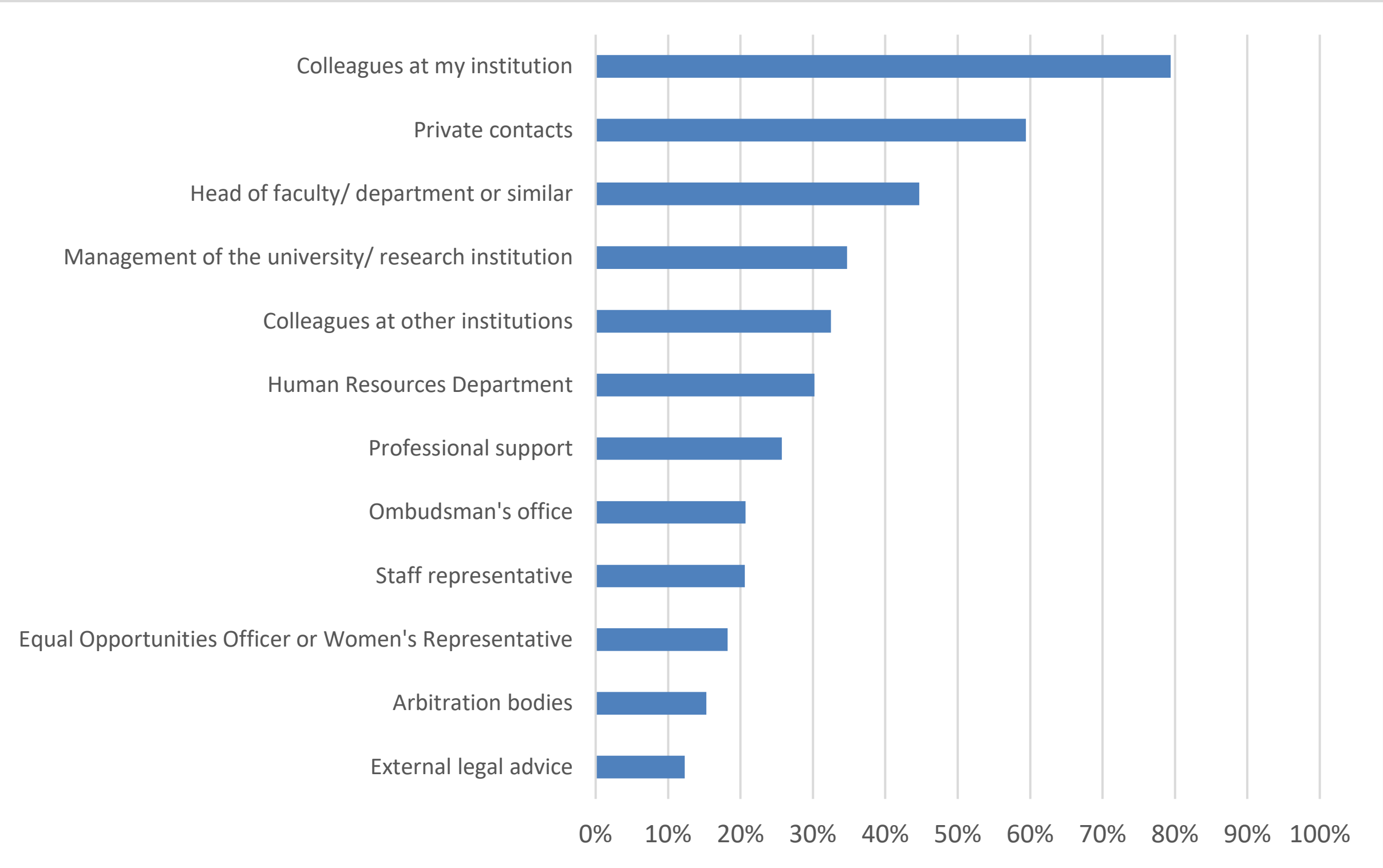

Source: DPG Survey 2024, own analyses, N= 251-255. *"In the event of conflict, other people or entities may provide support - or rather make things more difficult. How do you rate the support from the following people and entities?"*

**Table 3:** *Involvement of other actors in the conflict - illustrating quotes*

| **Category** | **Illustrating Quotes** |
|---|---|
| **Informal support** | • „The Ombudsperson made inquiries at the faculty if I really was abusing my power and bullying my coworkers. And some colleagues had heard such rumours and told the ombudsperson – there was no solidarity, no support for me, no one even thought it possible that there might be another side to it.“<br>• „I talked to my old mentor about it who gave me a lot of helpful advice.”<br>• „*[Helpful was]* a colleague, who had the same problem that I had.”<br>• „Other professors’ understanding – they had had similar experiences with that very person.”<br>• „Talking to my spouse about it helped me.” |
| **Institutional support** University or faculty/department leader(s) | • „I did not contact the university or faculty administration; the ombudsperson said that could backfire.”<br>• „I’m sure that the dean, the faculty, the department did try to do their best and be impartial. But I had the impression that it was very hard for them to be impartial, and there was no real solution or closure in the end either.” |
| Human resource department | • „I called the HR department *[for a termination during probationary period of a pre-doc]*. And they said basically: It’s too late. The board of staff representatives has already held their meeting, nothing doing.’ And effectively that meant: Now you have to live with *[this pre-doc]* for three years.”<br>• „I received a call from the Human Resource manager, who informed me about the official complaint. For data protection reasons, they were not allowed to disclose any details [...] The result was nights of speculation about who it could have been and what the reason might have been.” |
| Ombudsperson | • „I spoke to the ombudsperson and that was very helpful simply to calm me down, because I was angry, too. And the ombudsperson was comparatively – and calmed me down that such things happen.“<br>• „The ombudsman is in charge of scientific misconduct, but in my case, the problem wasn’t that – it was that my pre-doc thought they could successfully do a PhD - I thought so too in the beginning - but that was not the case at it turned out – and for such conflicts, no one really is in charge.”<br>• „The PhD ombudsperson presented the view of their client, but not a balanced view including my considerations.” |
| **Category** | **Illustrating Quotes** |
| Staff representative | • „At *[institution]*, there were two staff representatives - I didn’t think much of them. I just couldn’t believe they would really campaign for their colleagues.”<br>• „There’s a staff representative for scientific and non-scientific staff, and I call them for assistance, if necessary, because they don’t have anything to do directly with us and can neutrally approach the matter and listen to everyone.” |
| Equal Opportunity Office | • „[X] also went to the Equal Opportunity Office … and when I asked them, they basically refused to tell me anything about the complaint.”<br>• „There was an awesome Equal Opportunity Officer at *[university X]* – that would have been someone to contact.” |
| Arbitration bodies | • „Of the prescribed arbitration procedures, not one has helped.“<br>• „I would have needed some personal coaching before the arbitration process started.” |
| **External support** | • „All researchers involved in leading the group and a mediator were present at the discussion.”<br>• „There was an attempt to solve the conflict in a mediation process with a coach. A useful experience for me, however the coaching process did only provide a secluded environment to express the points of view of all parties. It did not lead to any changes.”<br>• „*[Helpful was]* the clear communication by the institution awarding the persons’ stipend.”<br>• „*[There was]* a meeting with a professional conflict mediator, mainly because in our institutes‘ management, there’s fear of conflicts.” |

Source: DPG Survey/Interviews 2024. Original quotes slightly smoothed and compressed for better readability and, if necessary, translated from German for the purpose of this table. For reasons of confidentiality, there are no pseudonyms or codes, and all references to genders, institution types etc. are obscured and paraphrased.

Private contacts and colleagues, both at the own and other institutions, were most frequently rated as helpful (figure 9), along with professional support and faculty management. The experiences with ombudspersons, external legal advice, arbitration bodies, staff representatives and the HR departments were mixed, while heads of the university or institutions and equal opportunity representatives were rated more often as a hindrance than as a help.

With respect to gender, we find only one difference – women evaluated equal opportunities' officers support more positively than men, although they were involved at an almost equal rate. Colleagues at the same institution were more frequently mentioned by physicists working with large-scale facilities who at the same time mentioned colleagues at other institutions the least frequently. Apart from that, we could not find any significant field differences.

**Figure 9:** *Evaluation of other actors' involvement*

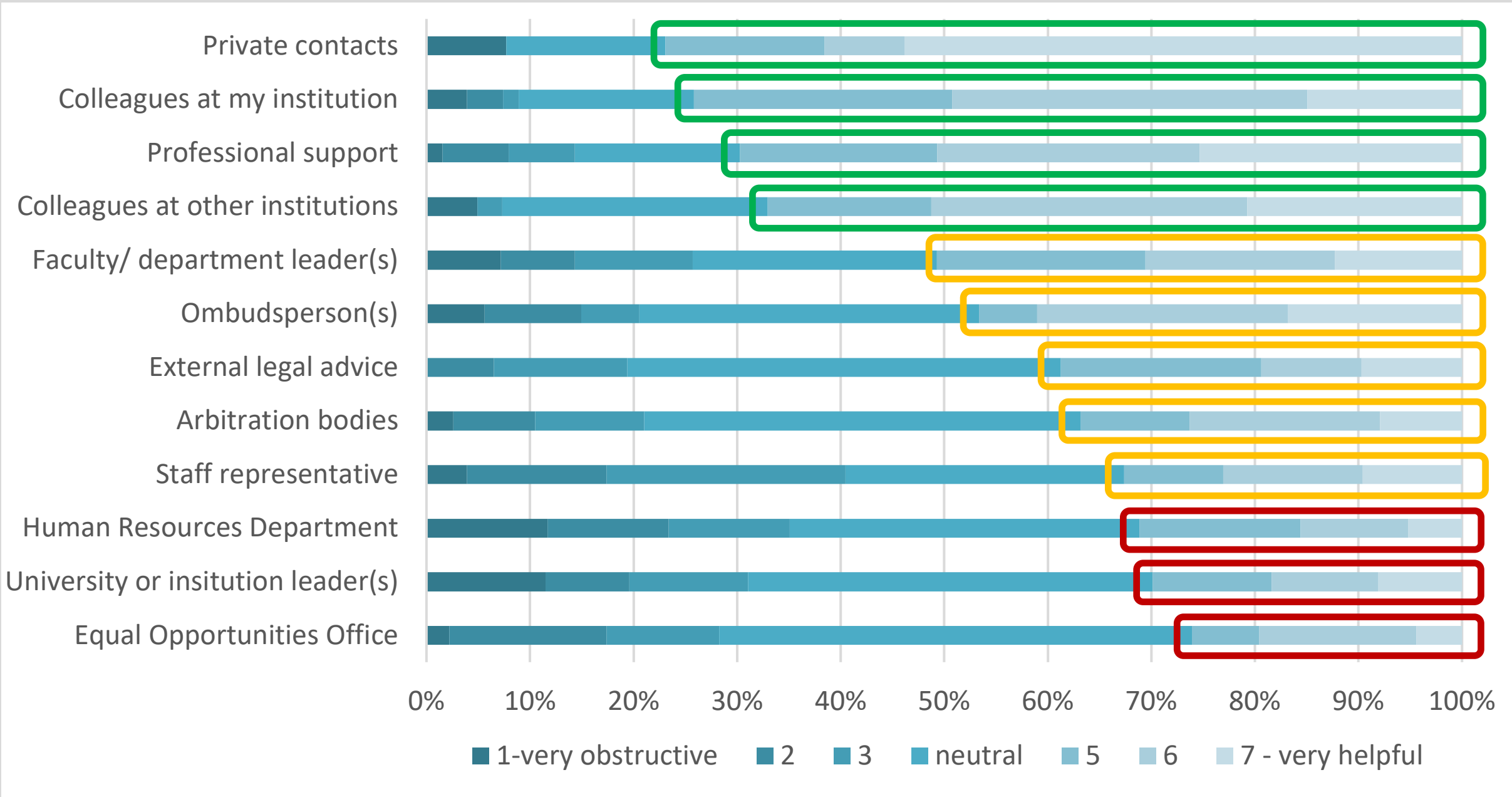

Source: DPG Survey 2024, own analyses, N= 31-201. *"In the event of conflict, other people or entities may provide support - or rather make things more difficult. How do you rate the support from the following people and entities?"*

## 4.4 Consequences of conflict – feared and real

Research leaders experienced a variety of worries due to conflict, most prominently they feared for the work climate (figure 10), with almost 80 % of research leaders mentioning it to at least some degree. More tangible worries concerned mainly the research itself, such delayed research, lower research quality and an inability to fulfil agreements with cooperation partners follow with a margin and were reported by less than half of research leaders. Adverse personal consequences to reputation and career or management sanctions were feared least with roughly 10 to 20 % mentioning these aspects.

**Figure 10:** *Feared consequences of conflicts*

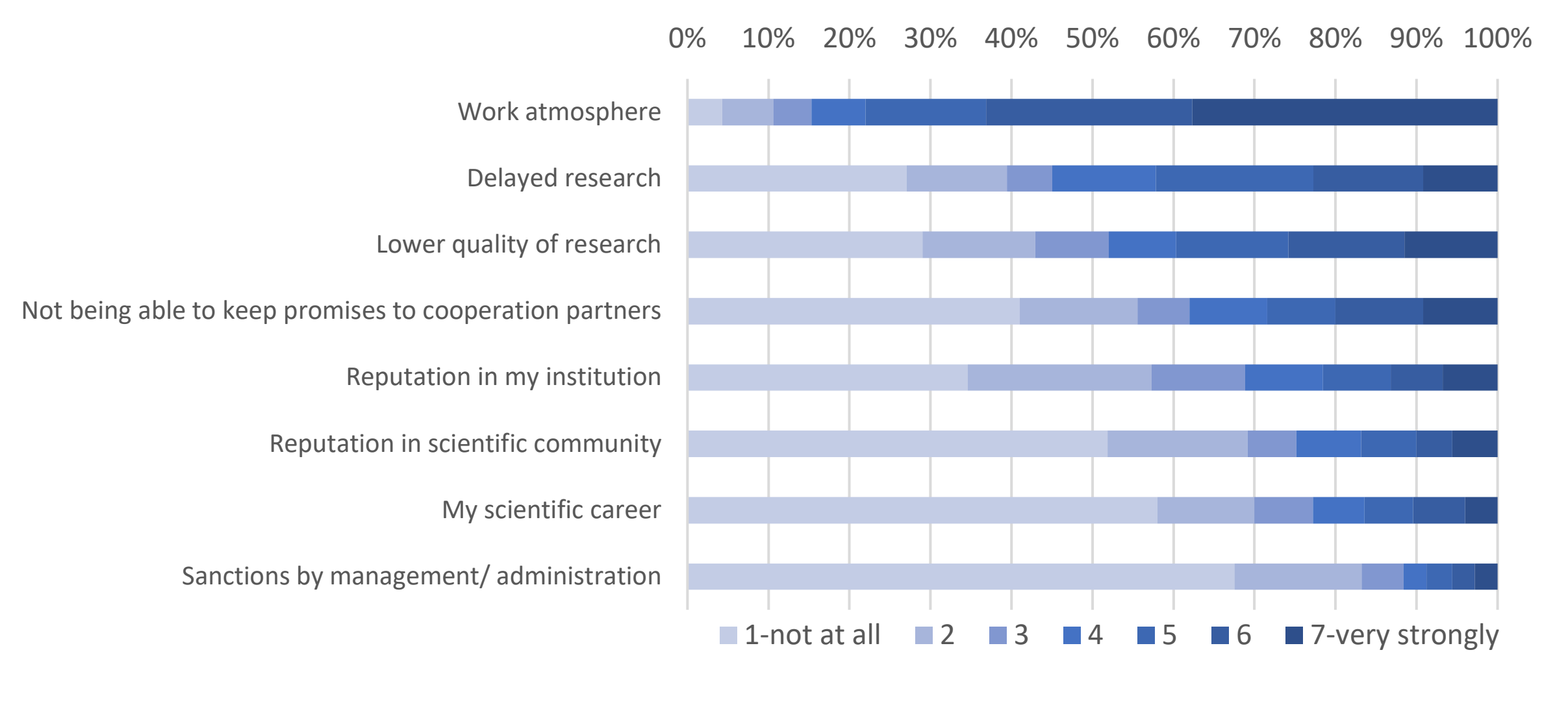

Source: DPG Survey 2024, own analyses, N= 255

Beyond the feared, we asked about the actual negative effects of the conflict (figure 11 and table 4). Research suffered most: leaders most frequently mentioned delays to at least some degree (44 %), followed by lowered quality (29.8 %) or results that could not be published (26.8 %). Direct personal damages, such as reputation loss, damage to one's own career and sanctions by management, only were reported in a minority of cases. No gender or field differences could be identified for feared or actual consequences.

**Table 4:** *Feared or real consequences of conflicts - illustrating quotes*

| Category | Illustrating Quotes |
|---|---|
| Impacts on research (Delay, lower quality, unpublished results) | • „It would have delayed everything – the post-doc, too, was on a fixed-term contract and needed the qualifications. So we let [X] be first author. We had the next project in the pipeline and wanted to get ahead."<br>• „For three years, I dragged that person more or less along […] the funds were gone and it was clear that there would be no progress in this area."<br>• „The conflict had negative consequences in the follow-up projects for me and my group – all was being negotiated behind my back. In the end, it contributed to the fact that we had to let our colleague go, also for financial reasons."<br>• „*[The conflict]* led to a lack of time which I would have rather spent supporting my team and doing better and more research."<br>• „[X] accused a colleague of using their results in a grant application without their consent. We could solve the conflict only by retracting the partial project concerned. The grant application wasn't successful in the end." |
| Impact on self (Reputation in my institution / in the community Scientific career Reputation in my community Sanctions) | • „With research leaders, social capital and reputation have such a high importance – and they can be damaged so easily – so of course, there are mutual dependencies."<br>• „My reputation was endangered – nor necessarily endangered, but potentially, and with me at the beginning of my career."<br>• „Many years later, *[the former conflicting party]* was asked by my chair to evaluate my performance as a supervisor and provided a negative report. This report was taken as the main reason for not promoting me to a better position." |

Source: DPG Survey/Interviews 2024. Original quotes slightly smoothed and compressed for better readability and, if necessary, translated from German for the purpose of this table. For reasons of confidentiality, there are no pseudonyms or codes, and all references to genders, institution types etc. are obscured and paraphrased.

**Figure 11:** *Actual consequences of conflicts*

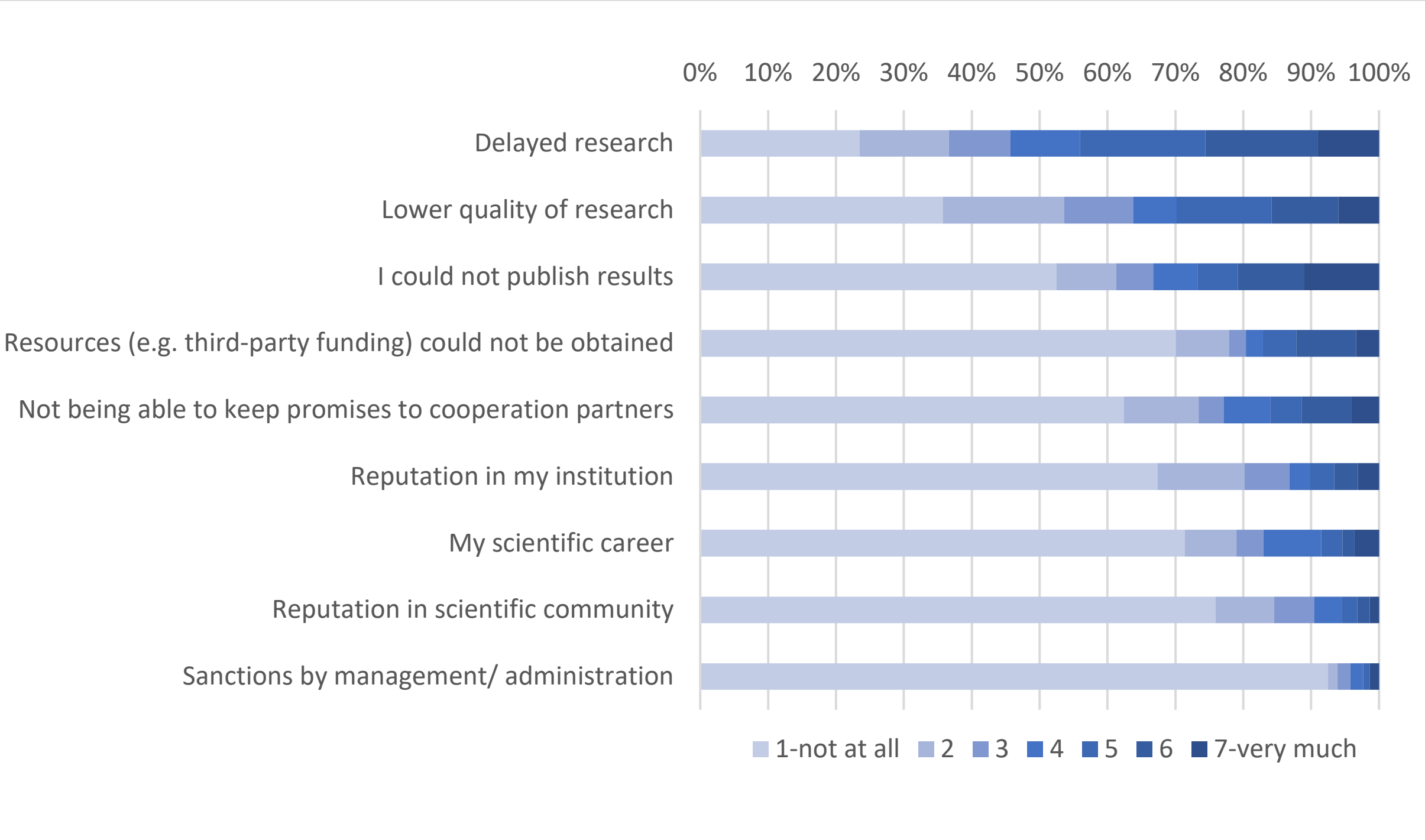

Source: DPG Survey 2024, own analyses, N= 254

When someone's doctoral or post-doctoral qualification was at stake (which was the case in about 63 % of reported conflicts), that person often could achieve their goal – in more than half of the cases even without changing their supervisor or institution (figure 12). In 8 % of cases, the academic qualification was not completed by the conflicting party. Probably, this failure was related to the conflict – either as consequence, cause or aggravating circumstance.

**Figure 12:** *Conflicts and qualification goals*

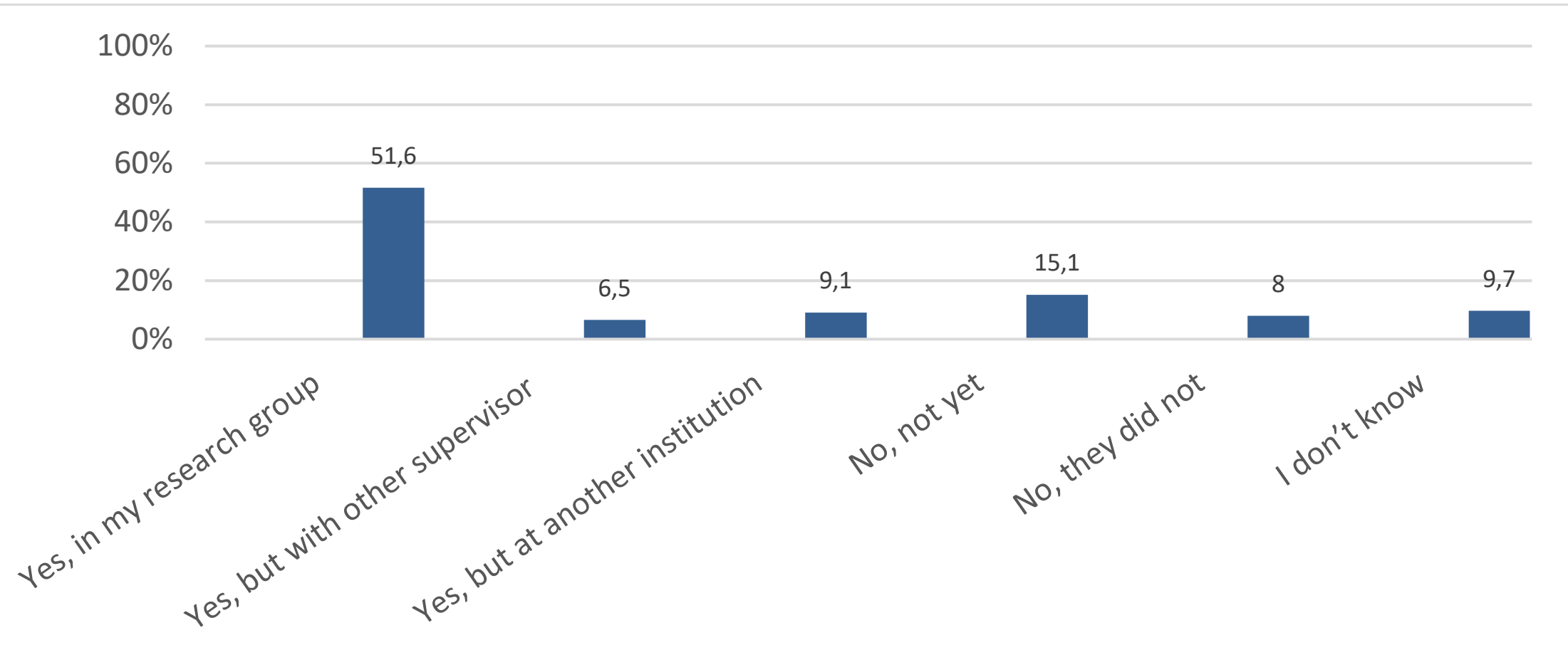

Source: DPG Survey 2024; own analyses; cases excluded where none of the main conflicting parties pursued a qualification goal, N= 186.*"If the main participant pursued a qualification goal: Was this goal achieved?"*

## 4.5 Personal responses by leaders to conflict experiences

Moderate to strong emotional responses to the reported conflict were widespread, above all anger and tension (figure 13), which were experienced by more than half of the research leaders at least to some degree. About a third reported disheartenment or depression as well as physical reactions to stress and strain. Doubts about one's own decision for research or the present organization did occur less frequently – in around 10 % of cases respectively at least to some degree.

**Figure 13:** *Personal emotional responses to conflicts*

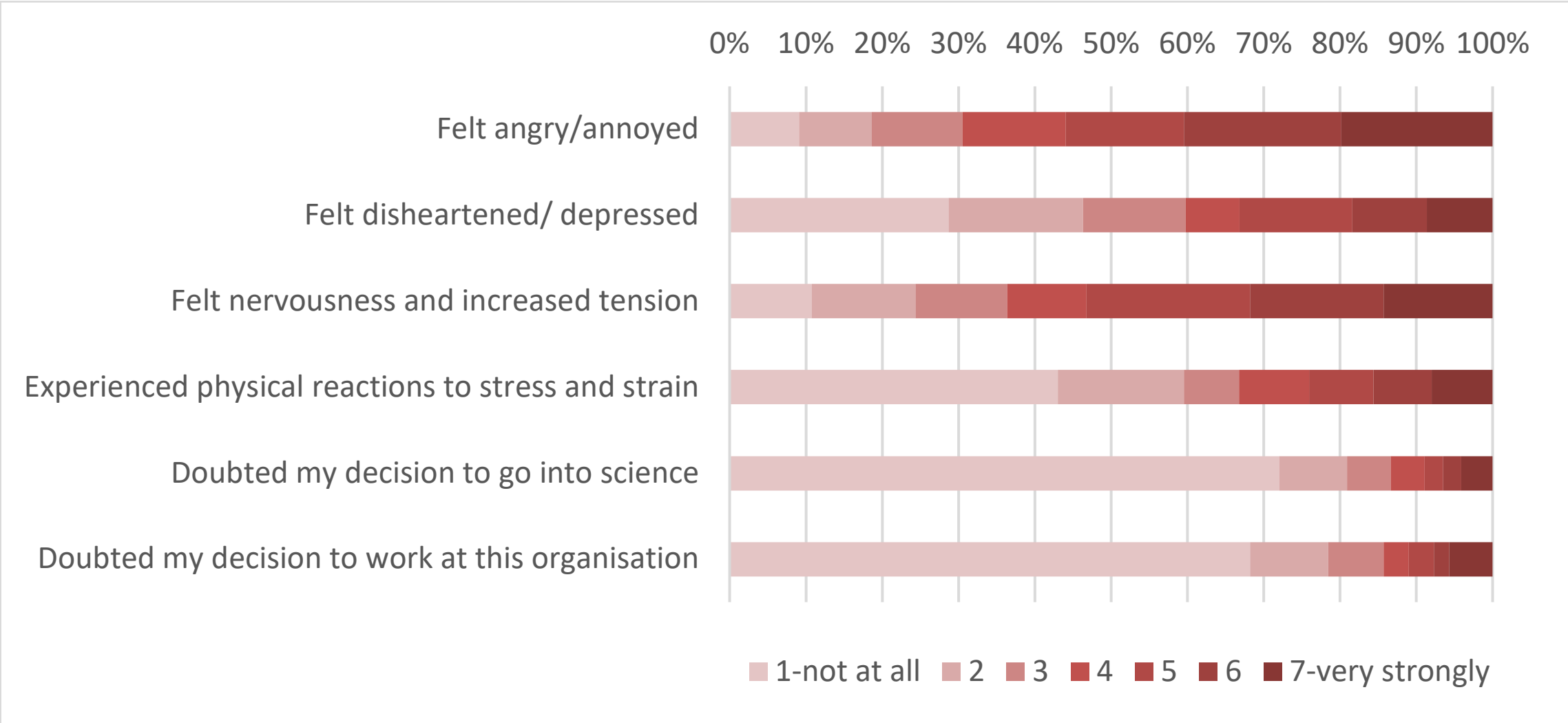

Source: DPG Survey 2024, own analyses, N= 252

In most cases (more than 90 %), researcher indeed responded to the experience by adapting their leadership behaviour (figure 14 and table 5). The most frequent measures concerned the way of organizing cooperation and communicating with team members: 55.7 % had started to communicate expectations more clearly, 52.2 % to improve meeting structures, and 52.2 % changed their way of interacting with individual research group members. 35.5 % amended their staff selection criteria, and about a quarter increased their professional distance. Steps to professionalize leadership by training or mentoring were undertaken by about a fifth. Also, a lowering of expectations (12.2 %) and changes in contract terms (5.9 %) were mentioned, as was downsizing of the research group (3.9 %).

**Figure 14:** *Changes in leadership style/behaviour in response to conflict experience*

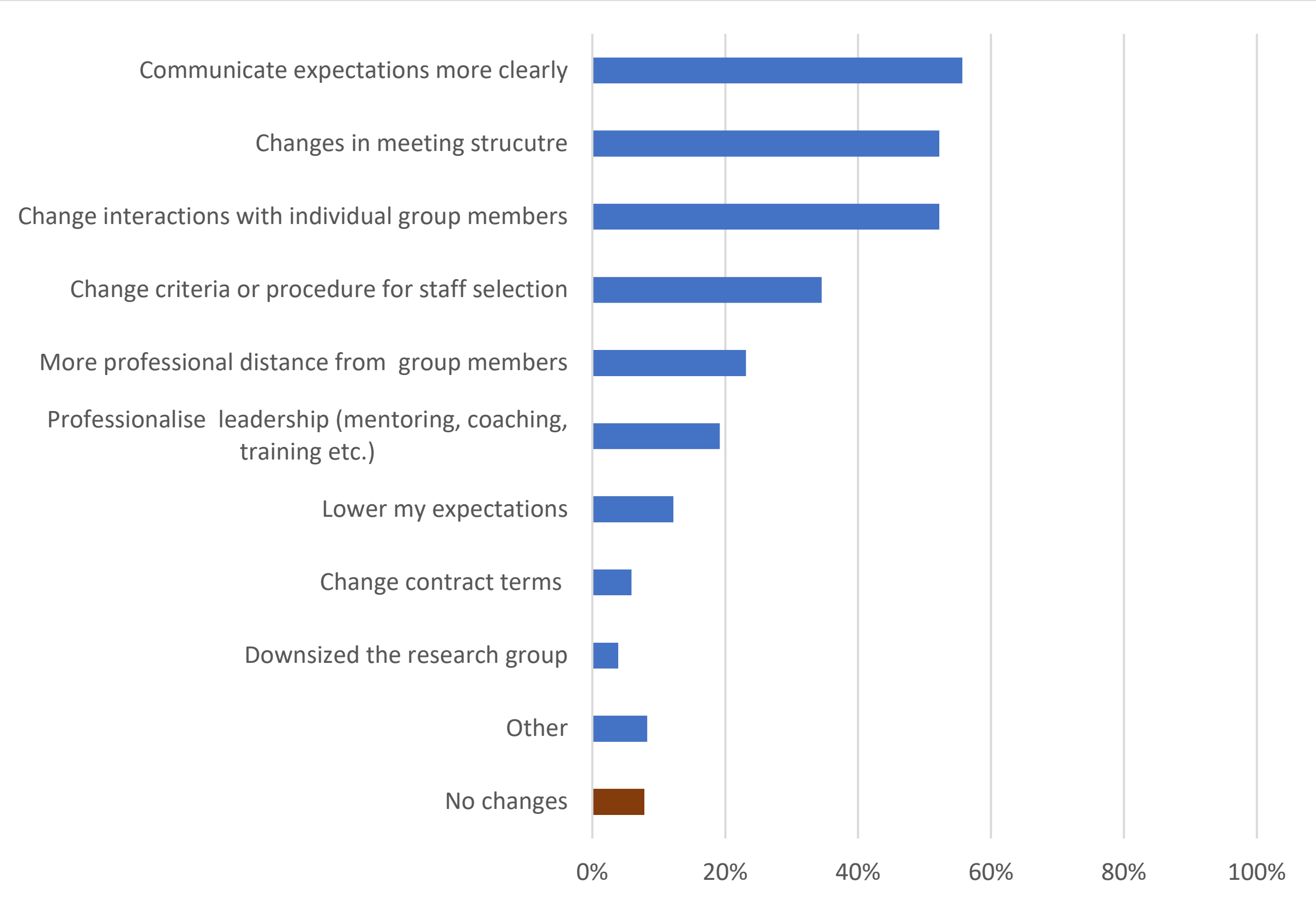

Source: DPG Survey 2024, own analyses, N=255

With regards to gender differences, female research leaders reported more emotional reactions throughout, and a slightly elevated tendency to doubt their decision for science and their organization. They also more often strived to professionalize their leadership through coaching or mentoring, and downsized their research groups. Field differences in changes in leadership behaviour concerned the adjustment of expectations, which occurred more frequently in large-scale facility research, and the changes in meeting structures, which occurred more often among theoretical physicists (not displayed). To complement the quantitative findings, table 5 presents illustrative excerpts on responses to conflict.

**Table 5:** *Responses to conflicts - illustrating quotes*

| **Category** | **Illustrating Quotes** |
|---|---|
| Organization of work | • „I've downsized my group, or rather, taken care not to enlarge it any further. My personal style really requires personal acquaintance, and that's not possible in a very large group."<br>• „I haven't lowered *[my expectations]* with respect to the quality of research, I think. But with respect to the amount a single individual can contribute – that expectation I'm handling more flexibly now and overall, yes, lowered it."<br>• „Holding obligatory meetings, planning work together, insisting that deadlines and agreements are kept."<br>• „I started to document every little meeting in a protocol so that my words can't be twisted around later - this effort is insane, especially if you consider that my whole career is based on verbal mutual agreement". |

| (Table 5, continued) **Category** | **Illustrating Quotes** |
|---|---|
| Formal aspects of re-cruitment | • „I am now only hiring doctoral students for one year at a time. This allows both sides to see if it is a good fit. And if that's not the case, people can quickly find something else if it doesn't work out. You could criticize that from the doctoral students' point of view, as it means they have no security from the outset. I understand that. But based on the experience I've had so far, I don't see any other option.”<br>• „There are so many really good candidates applying – and saying ‘no’ to those who aren’t a really perfect match, that’s a challenge.” |
| Communication | • „My way of communicating was completely without structure – I assaulted [X] on the corridor with something I wanted, while they would have needed like ‘I want to talk with you about this and that, when would it suit you?’ in order to be prepared for the talk.“<br>• „It is hard for me, but I’m explaining more why I do what I do. For example, I give my cramped schedule as a reason of interrupting coworkers as soon as I see that there’s nothing new since our last meeting.”<br>• „I can only expect from my team members what I express as an expectation - that was a learning process for me. Sometimes I thought, ‘It goes without saying that he or she will do this and that.’ And then it became clear that no, it's not obvious at all, and the person may not even know that it's important to me.” |
| Professionalize leader-ship | • „I personally have to consider what I can do to have or develop better or clearer leadership skills - I have a contact, it’s *[a professional coach]* – that’s my private thing, but I’ll look around at *[institution]* to see if they offer anything.”<br>• „I see this with others who start their own research groups – one has to realize that one’s now on a different plane and can’t really be ‘friends’ with one’s team members. I shrank back from making this clear, but that’s no good to anyone.”<br>• „After this conflict, I sought external counselling, and I took part in an extremely useful seminar on leadership and staff management by *[professional association]*. Since then, I can handle similar situations much better.“ |

Source: DPG Survey/Interviews 2024. Original quotes slightly smoothed and compressed for better readability and, if necessary, translated from German for the purpose of this table. For reasons of confidentiality, there are no pseudonyms or codes, and all references to genders, institution types etc. are obscured and paraphrased.

## 4.6 Evaluation of conflict development and conflict resolution

Regarding satisfaction with conflict development and its final resolution, ratings were all across the board and strikingly evenly spread from complete dissatisfaction to complete satisfaction (figure 15). Interestingly, the overall satisfaction with the resolution was higher, meaning that there were incidences, where the outcome was to the leaders’ satisfaction but not the way it was achieved, in spite of a sizeable correlation between the two ratings (Spearman’s rho=.67, p=.000). There were no gender differences, but with regard to fields, those with their own laboratories tended to be somewhat more satisfied with the development of the conflict, followed by theoretical physicists, while those doing large-scale facility research were least satisfied on average.

**Figure 15:** *Leaders' satisfaction with conflict development and resolution*

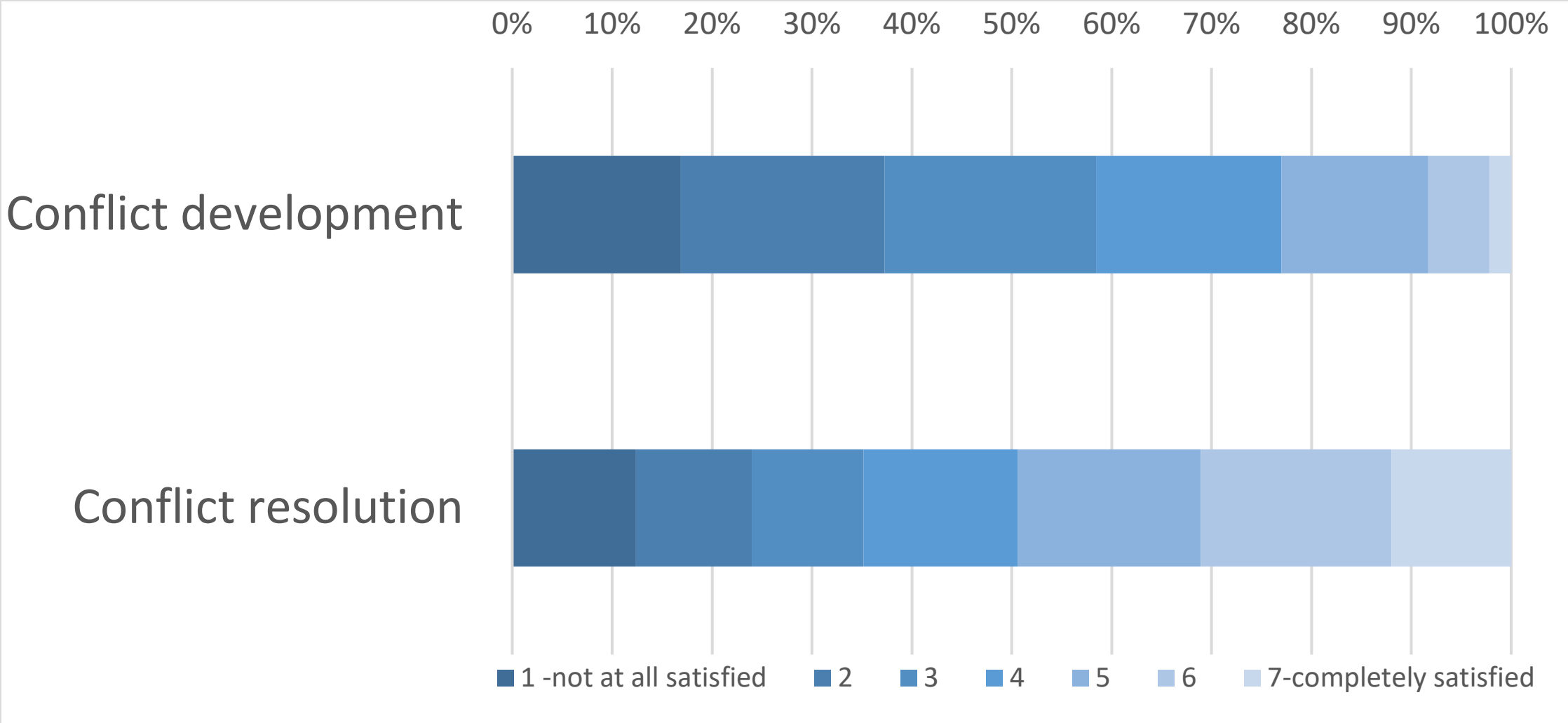

Source: DPG Survey 2024, own analyses, N= 231/251

# 5. Summary and outlook

## 5.1 Conflicts in research teams: Leaders' perspectives

This study was meant as a contribution to the understanding, prevention and management of conflict in research groups. We wanted to give research leaders an opportunity to describe their perspectives on workplace conflicts and evaluate the encompassing institutional reactions.

First, we looked at *experiences with complaints* (chapter 3). In general, research leaders who had made such an experience saw ample room at the top with respect to the fairness, transparency and constructive spirit of institutional handling of their situations. In this respect, the perception of the research leaders coincides with that reported in the survey studies based primarily on juniors' experience (see chapter 1.1). Since only about 12 % of our respondents reported such an experience, there were not enough cases to examine specific patterns in antecedents or consequences, such as differences between genders or fields.

We then focused more broadly on leaders' *experiences with conflicts* (chapter 4). As expected, conflict in research teams was a near-ubiquitous phenomenon, with less than 10 % of respondents who did not mention any conflict. The conflict themes were varied, and most widespread were authorship conflicts, lack of respect or noncollegial behaviours. This demonstrates that even in a highly rational research environment, interpersonal conflicts and communication difficulties are among the most common sources of conflicts.

In the next step, we examined *selected memorable conflicts*, without any restriction beyond leaders' subjective perception as noteworthy. The selected conflict themes correspond in their relative frequency to the ones mentioned most frequently for conflicts in general, supporting the notion that all kinds of conflict can become a noteworthy experience. Reported conflicts occurred between persons of all positions; but we found an unexpectedly large share of conflicts between postdocs, that is, among formal equals. In the present discussion, there is a focus on conflict between members in higher and lower positions where formal power is clearly unequal. However, formal hierarchies can also help reduce conflicts, and personal differences are most likely to occur or escalate in the absence of clear ranks and authorities for decision-making (Armstrong, 2012; Kühl & Schütz, 2022).

Most frequently *involved* (and *perceived as most helpful*) were colleagues at the same institution and private contacts – that is, informal sources of advice or support. Official institutional bodies were less often involved and often not perceived as helpful. This may be an effect of selectivity: Conflicting parties are more likely to seek official help in a particular complex or threatening conflict situation where some steps of escalation have already occurred and a satisfactory solution becomes more and more unlikely (Glasl, 1982; Scheppa-Lahyani & Zapf, 2023). In addition, such bodies don't have a general conflict resolution mandate but highly specialized roles and functions. Partly, they are even supposed to support a specific side - for example, employees in getting their rights (in the case of staff representatives) or female employees (in the case of equal opportunity's offices).

In the majority of the mentioned conflicts, there was *no serious harm done* to the research leaders involved, and qualification goals of conflicting parties could be reached. This is in line with our initial assumption that the majority of conflicts, even the memorable ones, do not escalate to an alarming degree. More widespread, however, were damages to research productivity, a good that probably interests both seniors and juniors both for their career advancement and their personal motivation and ideals and points to the importance of resolving conflicts adequately. The fact that only under a third of conflicts did not touch upon any qualification process serves as a reminder of how inextricably research is linked to qualification, and that conflicts often occur within this complex entanglement of collaborative knowledge production and certification of individual research performance.

Many research leaders reported to have gained *useful insights* from the conflict experiences, which prompted them to change and presumably improve their leadership behaviour with the aim to avoid similar experiences. This supports the notion that conflicts, even stressful and damaging ones, can be an opportunity to learn and grow and should therefore not be avoided and denied at all costs, but acknowledged as a fact of life and worked with.

*Satisfaction with conflict development and its final resolution* was fairly evenly distributed over the spectrum from complete dissatisfaction to complete satisfaction. Interestingly, there was a sizeable correlation between the two ratings, but the overall satisfaction with the resolution was higher than that with the process. This means that, in many instances, the outcome was to the leaders' satisfaction, but not the way it was achieved. This points to a sensitivity of research leaders to the importance of a good conflict resolution that does not equate personal "victory" with a satisfactory process.

*Gender differences* are scarce; they were most pronounced and conclusive with respect to emotional and practical responses: Females reported more negative feelings such as anger, sadness or doubt, more often sought professionalization for their own leadership role and adjusted research group sizes. In addition, the support of equal opportunities officers was rated as much more helpful by female research leaders. Within Physics, similarities are high between the subfields, although some *field-specific differences* point to subtle variations in the contexts for the development of conflicts, partly connected to different sizes of research groups.

## 5.2 Strengths and limitations

While providing valuable and new insights, our study has some limitations:

- *Selective participation and partly small n*: Both the interview and the survey sample are potentially biased. In all probability, the invitation more strongly attracted those to whom conflicts and their management are a personal concern – either because of their wish to become better leaders or because of their own (unpleasant) experiences. We do not have any means to estimate the degree of this bias or control for it. Moreover, while the number of participants is quite satisfactory overall, subgroups for some analyses are rather small. All generalizations, therefore, need to be undertaken with great caution and reasonable restraint.
- *Subjectivity:* Measures as perceived helpfulness of support or satisfaction with procedures may also be dependent how the respondents' own interests were achieved, and not only on objective justice or adequacy (if such a thing can be established). We generally have no means of making objective judgements in this study, but it should be borne in mind that these reports reflect subjective experiences.
- *One-sided perspective:* While adding to overall knowledge by focusing on the so-far underresearched leaders, we have no means to introduce two-sidedness on the level of individual conflicts, that is to say: we cannot take one conflict and juxtapose perceptions of all involved parties. A research design permitting this would be rather difficult to realize in an anonymous large-scale quantitative setting, but certainly of high relevance.
- *Universities or research organizations as conflict management context:* While these are of high relevance as formal context for conflict prevention and resolution in research, there are other influential actors (e.g. publishers, funding agencies or disciplinary professional associations). Such bodies could even be more relevant, since loyalty and identification of researchers with their university or department can be much lower than that with their research community or "science" itself, especially in times when professional self-government in universities is weakened (Schmidt & Richter, 2008; Janßen et al., 2021). The interplay between different actors should be taken more systematically into account in further studies.
- *Focus on researchers only:* By focusing on conflicts between research leaders and their pre- and postdoctoral team members, we could not assess the amount and kind of conflict with other individuals that research leaders have to lead as part of their team, such as secretaries or lab technicians. These, however, are instrumental for a research group's performance, and there are indications that they are also involved in conflictual workplace processes – maybe even to a higher degree than the academic staff (Striebing et. al., 2021; Schraudner et al., 2019). Further research should examine their perspectives and the dynamics of conflict between scientific and supportive staff more closely.

- *Descriptivity of the results:* While the findings paint a vivid picture of conflict experiences, we have not yet identified reliable patterns regarding different subgroups (such as gender or age). Also, it would be valuable to address possible recurring constellations of themes, actors and contextual factors that point to common or conflict setting by multivariate and/or data-driven methods (though clustering, or Latent Class Models). Such an approach could also be valuable for briefing leaders about “warning signs” to look out for in their groups and research contexts.

## 5.3 Useful lines of action

In the following, we will derive five potentially useful lessons and one important baseline we learned from this project as suggestions for people trying to improve the situation (figure 16 and table 6), relating them to relevant literature.

### Line of action 1: Normalize conflict

The research leaders in our study emphasize that there needs to be a more open acknowledgement in every part of the university or research organization that conflict and friction are to be expected in collaboration. This fits in with research on workplace processes (e.g. Zhao et al., 2019), which clearly shows that tensions and conflicts are very common and take up a sizeable portion of leaders’ time and attention. In addition, they point out that conflicts needn’t be seen, by their very nature, as dangerous and threatening, but rather as something that can and should be resolved with as little damage as possible, or even turned into a catalyst for improved collaboration. Normalization, in our participants’ view, also includes the admission that not every conflict can be turned into a win-win-situation, and that a resolved conflict does not necessarily mean that everybody will see the outcome as ideal.

### Line of action 2: Develop leaders’ professionality

While the research leaders firmly focus on research as their central and foremost task, they acknowledge a certain lack of awareness of potentially problematic team processes, common conflict constellations and leadership pitfalls both in themselves and in their colleagues. For leadership development programmes, they demand that leaders’ individual styles and priorities should be taken into account – a recommendation also present in previous research (e.g. Rehbock et al., 2023; Schmidt & Richter, 2008; 2009). Moreover, they recommend, again in line with the relevant literature, that training should happen when there is a change in professional roles and behaviours on the way from apprentice to colleague to leader (Braun et al., 2016; Laudel & Gläser, 2008; Rehbock et al., 2021). Another important issue mentioned is newer leaders’ need to understand what being a formal superior or superordinate in public organizations such as a university entails. Also, participants emphasize that the formats of training and coaching are crucial for acceptance and impact; in particular, a strong peer exchange component and reasonable demands on their scarce time resources are seen as indispensable. They acknowledge that voluntary measures such as training and coaching will likely draw a select crowd (e.g. Glatzel, 2023; Cheng & Zhu, 2025), but they point to the younger generation of professors being probably more open and willing to accept or even demand such support.

### Line of action 3: Develop awareness and active followership

Our study participants also mention the role of the *follower* as a relevant one for team collaboration success – an issue much less researched and conceptualized (e.g. Machovcova et al., 2023; Baitsch 2017). Becoming more knowledgeable and aware of their own roles and functions within research teams and the high interdependence of all work is recommended for pre- and postdoctoral researchers. Some leaders mention this also in connection with high international diversity of pre- and postdoctoral researchers, as well as generational differences, which can lead to followers' expectations and training not matching their leaders'. As a competent and active followership can even mitigate negative impacts from suboptimal leadership behaviour to a certain degree (May et al., 2014), supporting junior faculty in developing a follower identity compatible with the demands of the academic workplace can be an important building block in preventing conflicts from escalating and improving conflict management.

**Figure 16:** *Lines of action*

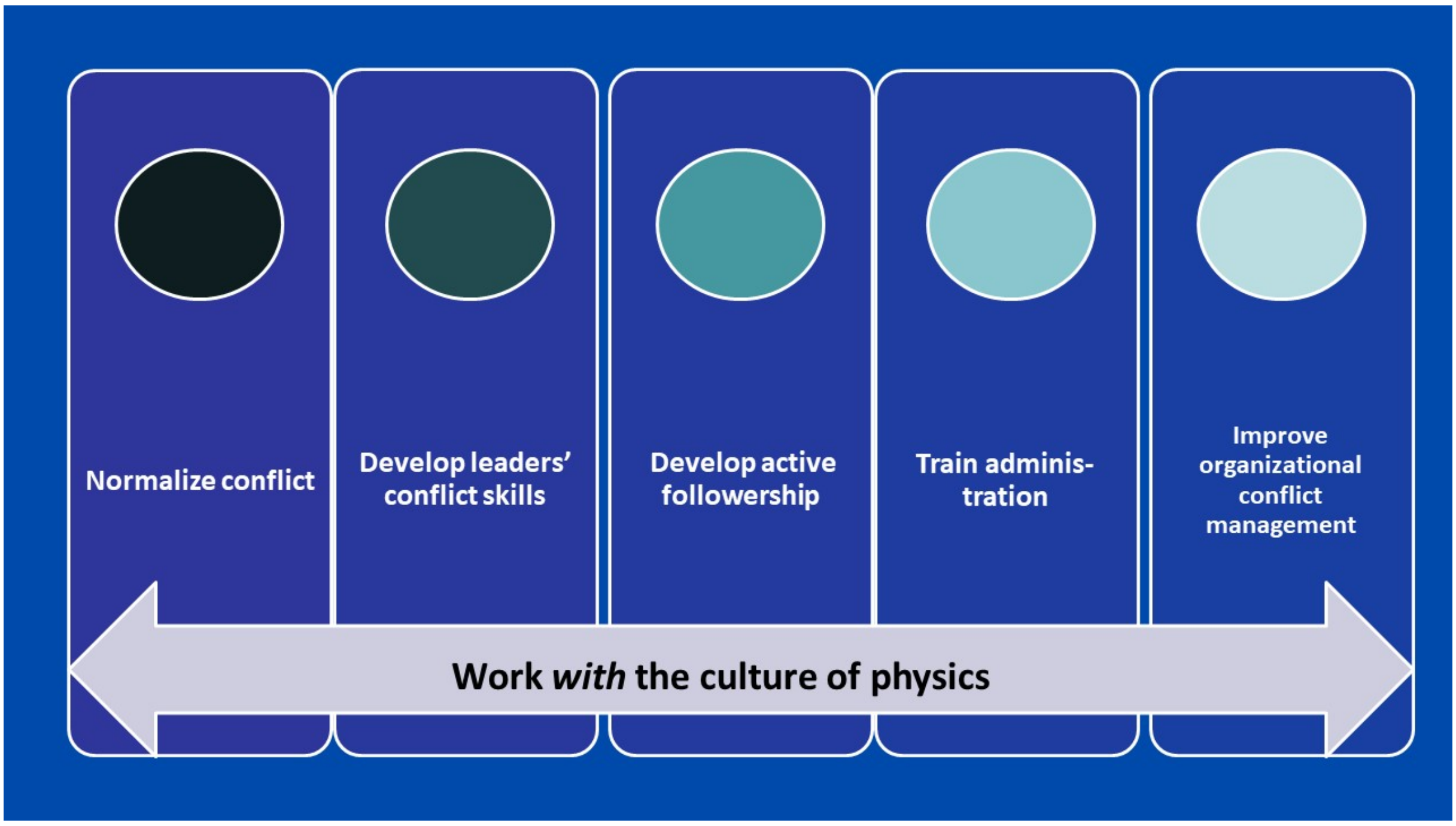

Source: DPG Interviews and Survey 2024, own analyses, N= 11/245

**Table 6:** *Lines of action - illustrating quotes*

| Category | Illustrating Quotes |
|---|---|
| Normalize conflict | • „After all, such situations *[rivalry between pre-docs]* are quite common.”<br>• „It’s important that one doesn’t let it grow into emotionalized conflicts. At present, I’m in a conflict with one of my colleagues, because they prioritize their tasks differently from how I see it…. Every six months or so we have a talk that we‘re both not really looking forward to, but we know by now that we can talk about it and find an agreement that allows us to carry on.”<br>• „There are always conflicts of interest, that’s in the very nature of research. Andy everyone thinks that they’re particularly conflict-averse whenever they have to talk to someone – but it’s never simple and pleasant.” |
| Develop leaders‘ (conflict) skills | • „… and we are not trained to deal with that!”<br>• „What I get from *[coaching]* is a kind of layer of abstraction I can apply to what happens to me, and then I can recognize und understand patterns and relations.“<br>• „*[Most programmes for leaders]* are not aligned with the need for efficiency and the timetables of us research leaders. Information density is too low, exchange density is too low, and way too much sitting and looking at slides.”<br>• „The scientific aspect of the work is not separable from the interpersonal communication. The latter part requires a lot of attention when working in groups. Scientist are often not well trained in this field, and they are rather encouraged to misbehave in order to advance in their career.” |
| Develop active followership | • „*[Predoctoral researchers]* are not helpless or powerless. I think they hardly realize that.“<br>• „*[Junior researchers*] are grown-ups, they can decide for themselves and… should take responsibility for speaking up – ‘What are my basic parameters?’ And say what they need to make this work”.<br>• „I miss a contact person, or obligatory coaching, to prevent students from informing themselves in questionable internet fora about their rights and duties, ending up with false expectations.” |
| Train administration | • „The people leading HR departments ought to be trained and have explained to them how universities work.“<br>• „People from private enterprises take over a leadership position in the university’s finance or human resource department, and lead the way they know it. Partly that works very well and introduces structure into formerly chaotic systems, but often it doesn’t, really.”<br>• „The less administration in departments puts obstacles in the researcher’s way, the less difficult it is to work and to focus on what it should be about - research, that is, not constant discussions about contracts, rooms, equipment etc.” |
| Improve organizational conflict management | • „There are no transparent procedures, everyone has his or her circle of friends and stakeholders. […] And I’d rather get help from outside, from people I can be sure that they don’t have their own agenda in the matter.”<br>• „At *[institution]* I really appreciate that they take conflict management seriously – there are manuals or guidelines for leaders, and webpages with these models of escalation and de-escalation that have been implemented.”<br>• „In that case, it was great that we had to do these regular *Mitarbeitergespräche*, because I said ‘It’s the management, they want us to have this talk, with two weeks for preparation, and this catalogue of questions’ – most of them were completely irrelevant but some were just the crucial ones, so that worked excellently in that case.“ |
| Work with the culture of Physics | • „Training staff need to be attuned to the special groups. They can have the same course for people in humanities and science but with a sensibility for field-specific differences – how groups are organized and such.”<br>• „My impression is that there’s a trend to transplant instrument, methods, structures from the private sector into academia, which I find dangerous and counterproductive.” |

Source: DPG Survey/Interviews 2024. Original quotes slightly smoothed and compressed for better readability and, if necessary, translated from German for the purpose of this table. For reasons of confidentiality, there are no pseudonyms or codes, and all references to gender, institution types etc. are obscured and paraphrased.

### Line of action 4: Support administration in working together with researchers

Study participants mention that involvement with management or administrative units is often full of friction, and that this contributes to the generation of conflict as well as suboptimal or escalating conflict processes. These experiences and perceptions reflect the dual nature of universities, which encompass two organizational cultures under the same roof - one characterized by collegial self-governance of research experts, the other by a formal bureaucratic logic of public administration (Banscherus, 2018; Nickel, 2012). Incompatibilities between the two cultures and their goals and rules have been identified in the literature as a frequent aggravating factor in conflicts (Baitsch, 2017; Jungwirth, 2025; Nickel, 2012). While not suggesting any concrete training measures, our study participants express a desire to have administration being better attuned to the needs of researchers and research, and to seeing themselves more as enablers of the universities core functions. Whenever research leaders report instances of constructive experiences, they show intense appreciation and pleasure, indicating that there is a lot to be gained by fruitful cooperation between research and administration.

### Line of action 5: Improve organizational procedures

Research leaders in this study voice some degree of frustration about intransparent procedures, and express reservations about the helpfulness of their organizations' designated conflict management actors such as ombudspersons, staff representatives and equal opportunity offices. These actors' circumscribed jurisdictions seem frequently inapplicable to their situations, or leaders perceive them as not acting equally for their benefit or an equitable solution. Again, participants mentioned positive experiences with a high degree of appreciation, relief and respect for the skills of the involved persons, highlighting their own vulnerability to suboptimal conflict management. Conflict management has indeed been shown to make up a large part of the duties of ombudspersons, staff representatives and equal opportunity offices, but beyond these bodies, there are seldom any transparent and explicit organizational forms (Hochmuth, 2014: Hoormann & Matheis, 2012). While our participants see the institutions responsible for creating a good conflict management environment, they also emphasize the need to include neutral outside actors for a competent but unbiased dealing, such as external arbitrators, mediators or coaches, as pointed out also in relevant studies (e.g. von Knobelsdorff, 2025; Hochmuth, 2014).

### Bottom line: Work with the culture of science

Our participants express an adamant critique of management or administration methods and procedures that are imported from other contexts (namely, businesses as private enterprises or other public organizations), which, in their view, are inappropriate and not helpful for what they see as the quintessential purpose of a research organization. This is a common perception of researchers (e.g. Baitsch, 2017; Hochmuth, 2014) which – be it justified or not – represents a serious obstacle to managerial approaches to the improvement of governance and management at universities. Re-shaping conflict perception and handling in all parts of a university or research organization amounts to nothing more or less than a cultural shift (Hochmuth 2014).

Only if researchers are involved in a meaningful way in this shift and see that it benefits their research and their disciplines, profound and sustainable change can be achieved (Jungwirth, 2025; Straatmann et al., 2023). At the same time, this involvement must not interfere unduly with what researchers can do best and want to do most, that is: research.

# Appendix

## **Table 7:** *Characteristics of respondents – details by gender*

| Area | Theoretical Physics | Experimental Physics – own laboratory | Experimental Physics – Large Equipment Research | Other |
|---|---|---|---|---|
| *Female* | *22,1%* | *48,5 %* | *19,1%* | *10,3%* |
| *Male* | *22,8%* | *45,5%* | *17,5%* | *14,2%* |
| N | 63 | 129 | 50 | 37 |
| Current Position | Professor (including Junior professorships with or without tenure, visiting, honorary or senior professors and emeriti) | postdoctoral research associate | Other | |
| *Female* | *61,8%* | *33,8 %* | *4,4 %* | |
| *Male* | *58,4%* | *33,0 %* | *8,6 %* | |
| N | 164 | 92 | 21 | |
| Current organization | University | Research Organization | Other | |
| *Female* | *61,8%* | *27,9%* | *10,3%* | |
| *Male* | *57,1%* | *28,3%* | *14,6%* | |
| N | 163 | 79 | 32 | |
| Country | Germany | Other | | |
| *Female* | *79,4%* | *20,6%* | | |
| *Male* | *81,9%* | *18,1%* | | |
| N | 226 | 52 | | |
| Nationality | German | Other | | |
| *Female* | *81,5%* | *18,5%* | | |
| *Male* | *90,9%* | *9,1%* | | |
| N | 242 | 31 | | |

*Source: DPG Survey 2024, own analyses. Only persons included who reported their gender; distributions and numbers therefore may vary between tables and figures in this text*